\documentclass[prd,floatfix,nofootinbib,showpacs,showkeys,superscriptaddress]{revtex4}
\usepackage{exscale}                  % correct scaling of math-symbols
\usepackage[intlimits]{amsmath}       % improved mathematical typesetting
\usepackage{amsfonts}
\usepackage{amssymb,amscd}
\usepackage[dvips]{epsfig}                   
\usepackage{array}
\usepackage{wrapfig}
\usepackage{graphicx,subfigure}
\newcommand{\beqa}{\begin{eqnarray}}
\newcommand{\eeqa}{\end{eqnarray}}
\newcommand{\beq}{\begin{equation}}
\newcommand{\eeq}{\end{equation}}
\newcommand{\gS}[1]{#1\!\!\!\!\!\not~}	
\newcommand{\GS}[1]{#1\!\!\!\!\!\!\!\not~}	

\newcommand{\qslash}{\gS{q}}
\newcommand{\pslash}{\gS{p}}
\newcommand{\Pslash}{\GS{P}}

\newcommand{\intq}{\int\!\!\frac{d^4q}{(2 \pi)^4}}

\begin{document}

\large

\title{
 \hspace*{\fill}{\small\sf MIT-CTP 3964}\\[4mm]
 On Gribov's supercriticality picture of quark confinement}
\author{Christian~S.~Fischer}
\affiliation{Institute for Nuclear Physics, 
 Darmstadt University of Technology, 
 Schlossgartenstra{\ss}e 9, 64289 Darmstadt, Germany}
\affiliation{GSI 
Helmholtzzentrum f\"ur Schwerionenforschung GmbH, 
  Planckstr. 1,  64291 Darmstadt, Germany.}
\author{Dominik Nickel}
\affiliation{Center for Theoretical Physics,
  Massachusetts Institute of Technology,
  Cambridge, MA 02139, USA}
\author{Richard Williams}
\affiliation{Institute for Nuclear Physics, 
 Darmstadt University of Technology, 
 Schlossgartenstra{\ss}e 9, 64289  Darmstadt, Germany}
\date{\today}

\begin{abstract}
Some years ago Gribov developed the so-called supercritical
light quark confinement scenario. Based on physical arguments
he conjectured a drastic change in the analytical properties 
of the quark propagator when the back-reaction of Goldstone bosons
(pions) is considered. We investigate this scenario and provide 
numerical solutions for the quark propagator in the complex plane
with and without the pion back-reaction. We find no evidence for 
the scenario Gribov advocated. As an aside we present a novel 
method to solve the quark Dyson-Schwinger equation in the complex 
plane and discuss new characteristics of dynamical chiral symmetry 
breaking in our truncation scheme.
\end{abstract}

\pacs{12.38.Aw, 12.38.Gc, 12.38.Lg, 14.65.Bt}
\keywords{Quark propagator, dynamical chiral symmetry breaking, 
light mesons}

\maketitle
%%%%%%%%%%%%%%%%%%%%%%%%%%%%%%%%%%%%%%%%%%%%%%%%%%%%%%%%%%%%%%%

\section{Introduction}\label{sec:intro}

The phenomenon of confinement is usually thought of as originating
in the Yang-Mills sector of QCD. In the quenched theory with heavy
sources confinement thus understood manifests itself in the behaviour
of the Wilson loop at large distances; here an area law is associated
with flux-tube formation of colour-electric fields and a linear rising 
potential for heavy quarks. In the full theory,
however, the colour-electric string between these charges breaks due to
the creation of light quark-antiquark pairs. Therefore the potential
is no longer rising but levels out at large distances. Thus in a sense, 
made precise \emph{e.g.} in \cite{Greensite:2003bk,Greensite:2006sm}, full 
QCD is not confining.

Nevertheless, free colour-charges are absent in the real world and the
precise mechanism for this absence has to be determined in full QCD
with realistic quark masses. This so-called colour-confinement mechanism
is still elusive even after three decades of intense efforts. In a series 
of (partly unfinished) papers \cite{Gribov:1999ui,Gribov:1992tr,Gribov:1998kb} 
Gribov developed a scenario of quark confinement arising from the 
supercriticality of colour charges. The basic idea, summarised 
in \cite{Dokshitzer:2004ie,Ewerz:2006vw}, is the binding
of quarks with positive kinetic energy 
within a bound state of total negative energy. In order to guarantee 
a stable vacuum these states have to be filled up, therefore enforcing a 
vacuum with occupied quark states of positive kinetic energy
in addition to the
negative energy quark states of the conventional Dirac sea. Consequently, the
Pauli principle prevents single quarks from propagating and there can be
no corresponding asymptotic states of single quarks.

According to Gribov \cite{Gribov:1999ui}, an essential ingredient in this 
picture is the appearance of Goldstone bosons due to the dynamical breaking 
of chiral symmetry. The Goldstone bosons, identified with the pseudoscalar
pions, he conjectured to change the analytical structure of the quark 
propagator in such a way that the resulting quarks are confined by the 
supercritical mechanism. 
It is the purpose of this paper to critically investigate the actual influence
of Goldstone bosons on this structure.

To this end we employ a truncation scheme for the quark Dyson-Schwinger
equation (DSE) and the quark-gluon vertex DSE developed in ref.~\cite{Fischer:2007ze},
which leads to a quark self-energy governed by non-perturbative gluon and
pion exchange. The structure of the resulting DSE for the quark propagator
is similar to the equation Gribov 
% considered 
started with
originally.
% However, we managed 
% to avoid some of the approximations Gribov was forced to make in order to be
% able to solve his DSEs.
% In this, to our mind improved, truncation scheme we 
% obtain information on the analytic structure of the quark propagator with 
% and without pion back-reaction by a combination of several methods.
In contrast to Gribov, we work with the DSEs as coupled integral
equations rather than their -- in principal equivalent -- differential
formulation that he favoured for analytical studies.  However, in
converting an integral equation into a tractable differential equation many
approximations must be employed.  
%%% RW: Do we want to put the detail and contrast of the truncations here?
Instead we work directly with the integral equation 
and apply a truncation scheme that contains the same features
implemented by Gribov, which has been used as a basis for hadron phenomenology 
and comparisons to lattice QCD results.  In this
% , to our mind improved,
truncation scheme we 
obtain information on the analytic structure of the quark propagator with 
and without pion back-reaction by a combination of several methods. 
As a result we find no evidence in favour of Gribov's conjecture.

The paper is organised as follows. In section two we outline our
truncation for the gluonic part of the quark-DSE, together with our
approximation scheme for the hadronic part of the vertex, following the
procedure of~\cite{Fischer:2007ze}. We specify our method for exploring
the analytic structure of the quark-propagator in the complex plane, leaving
details of the implementation to appendix~\ref{complexplane}. In
section three we present our numerical results, and finally give our
conclusions.

\section{The approximation scheme for the quark-DSE\label{interaction}}

\subsection{Gluon exchange part}

\begin{figure}[ht]
\centerline{\epsfig{file=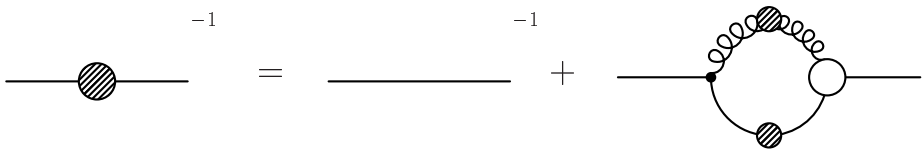,width=12cm}}
\caption{The Schwinger-Dyson equation for the fully dressed
quark propagator}\label{fig:dse1}
\end{figure}

The full Dyson-Schwinger equation for the quark propagator is displayed
diagrammatically in fig.~\ref{fig:dse1}. 

\noindent With the dressed inverse quark propagator $S^{-1}(p)=i{\pslash} A(p^2) + B(p^2)$ 
and its bare counterpart $S^{-1}_{0}(p) = i \pslash + m$ the equation
is given by
\begin{eqnarray}
  S^{-1}(p) &=& Z_{2}S^{-1}_{0}(p)
  + g^{2}C_{F}Z_{1F}\intq\,  \gamma_{\mu}S(q)\Gamma_{\nu}(q,k)
  D_{\mu\nu}(k)   \,, \label{DSE1}
\end{eqnarray}
with $k=p-q$, the Casimir $C_{F}=(N_{c}^{2}-1)/(2N_{c})$ and 
the renormalization factors $Z_{1F}$ of the quark gluon vertex and
$Z_2$ of the quark propagator.
In the course of this work we will consider the case $N_f=2$ and $N_c = 3$.
The vector and scalar dressing functions $A(p^2)$ and $B(p^2)$
can be recombined into the quark mass 
$M(p^2) = B(p^2)/A(p^2)$ and the quark wave-function 
$Z_f(p^2) = 1/A(p^2)$. The dressing functions depend on the fully 
dressed quark-gluon vertex $\Gamma_{\nu}(q,k)$ and the gluon propagator 
\begin{equation}
D_{\mu\nu}(k) = \left(\delta_{\mu \nu} 
- \frac{k_\mu k_\nu}{k^2}\right)                                        
\frac{Z(k^2)}{k^2} = P_{\mu \nu} \frac{Z(k^2)}{k^2}\,, \label{gluon}
\end{equation}
with the gluon dressing function $Z(k^2)$. Up to, 
for our purposes minor,
details in the far infrared the function $Z(k^2)$ is well known from 
both lattice calculations and Dyson-Schwinger equations (for a review
see \emph{e.g.} \cite{Fischer:2006ub}). 

Throughout this paper we will work in the Landau gauge as opposed to the
choice of Feynman gauge Gribov adopted in his work. We expect that if the
physical mechanism for quark confinement is triggered by the back-coupling of
Goldstone
bosons to the quarks this mechanism should be present in all continuously
connected gauges. We prefer
Landau gauge over Feynman gauge because the gluon dressing function is well
known there (see below) and the tensor structure of the propagator is
particularly simple\footnote{In other linear covariant gauges the propagator 
is given by $D_{\mu\nu}(k) = 
\left(\delta_{\mu \nu} - \frac{k_\mu k_\nu}{k^2}\right)
\frac{Z(k^2)}{k^2} + \zeta\frac{k_\mu k_\nu}{k^4}$, where $\zeta$
is the gauge parameter and $\zeta=1$ for Feynman gauge. Due to an exact
Slavnov-Taylor identity the longitudinal part of this propagator remains 
undressed. Only for a nonlinear gauge condition can the propagator
be rewritten as $D_{\mu\nu}(k) = \frac{\delta_{\mu \nu}}{k^2} \alpha(k^2)$
\cite{Gribov:1998kb,Ewerz:2000qba}, where the choice of the running coupling
as the dressing function is insprired from the Abelian theory. In the
non-Abelian
case such a choice already represents a combination of dressings for the gluon 
propagator and the quark-gluon vertex.}.

The other input into eq.~(\ref{DSE1}) is the fully dressed quark-gluon
vertex $\Gamma_{\nu}(q,k)$. An approximation for the vertex in terms
of the quark self-energies and the dressing function $G(p^2)$ of the
propagator of Faddeev-Popov ghosts has been developed in 
\cite{Fischer:2003rp,Alkofer:2003jj}.
The ansatz
\begin{equation}
  \Gamma_\nu(q,p) = V_\nu^{abel}(q,p) \, 
                        W^{\neg abel}(q,p) \;,
  \label{vertex}
\end{equation}
with
\begin{eqnarray} 
  W^{\neg abel}(q,p) &=& G^2((q-p)^2) \:\tilde{Z}_3 \; ,
  \nonumber
\\
  V^{abel}_\nu(q,p) &=& \Gamma_\nu^{CP}(q,p) \nonumber \\ &=&
  \frac{A(p^2)+A(q^2)}{2} \gamma_\nu + i
  \frac{B(p^2)-B(q^2)}{p^2-q^2} (p+q)_\nu
  \nonumber\\
  && {} + \frac{A(p^2)-A(q^2)}{2(p^2-q^2)}
  (\pslash+\qslash)(p+q)_\nu
  \nonumber\\
  && {} + \frac{A(p^2)-A(q^2)}{2} \left[(p^2-q^2)\gamma_\nu -
    (\pslash-\qslash)(p+q)_\nu \right] \nonumber\\
  && \hspace*{3.5cm} \times
  \frac{p^2+q^2}{(p^2-q^2)^2+(M^2(p^2)+M^2(q^2))^2} \;,
  \label{CP}
\end{eqnarray} 
where the ghost wave-function renormalisation $\tilde{Z}_3$
has been shown to lead to a quark-DSE which has the correct ultraviolet
asymptotic limit and respects multiplicative renormalisability. In addition,
the Abelian part $V^{abel}_\nu$ of the construction is identical with the 
so-called Curtis-Pennington vertex $\Gamma_\nu^{CP}$ \cite{Curtis:1990zs}. 
Its first three terms have been shown by Ball and Chiu \cite{Ball:1980ay}
to satisfy the Abelian Ward-Takahashi identity (WTI), 
\begin{equation}
i k_\nu \: \Gamma_\nu^{QED}(q,p) = S^{-1}(p) -  S^{-1}(q),
\label{WTI}
\end{equation}
with the quark momenta $q$ and $p$ and the gluon momentum $k=q-p$. As found in 
\cite{Alkofer:2003jj}, the presence or absence of the scalar interaction term 
proportional to $(p+q)_\nu$ is of particular importance for the analytical 
structure of the quark propagator. In section \ref{results} we therefore 
contrast results obtained with the ansatz (\ref{CP}) also with the 
simpler vertex 
\begin{eqnarray}
  W^{\neg abel}(q,p) &=& G^2((q-p)^2) \:\tilde{Z}_3 \; , \nonumber\\
  V^{abel}_\nu(q,p)  &=&  \frac{A(p^2)+A(q^2)}{2} \gamma_\nu\,, \label{1BC}
\end{eqnarray}
which does not contain the
scalar interaction term. Since the approximation (\ref{1BC}) consists of only the
first term of the Ball-Chiu solution of the WTI we will refer to it as 
`1BC-vertex'. It represents a form of the rainbow-ladder approximation 
of the quark-DSE which has been successfully applied to the physics of light 
mesons \cite{Fischer:2005en}. 

We wish to emphasize, that the ansatz (\ref{vertex}) for the quark-gluon vertex
has similar properties as a recent explicit solution of the quark-gluon vertex 
DSE \cite{Alkofer:2006gz}. In particular the infrared singularity 
$W^{\neg abel}(q,p) \sim G((q-p)^2) \sim ((q-p)^2)^{-2\kappa}$ with 
$\kappa \simeq 0.595$ \cite{Zwanziger:2001kw,Lerche:2002ep} present in all 
tensor structures of the vertex is also an approximate property of the 
explicit solution which is proportional to $((q-p)^2)^{-1/2-\kappa}$ 
\cite{Alkofer:2006gz}. We need to keep in mind, however, that the relative 
strength of the different tensor structures in the full vertex may not  
be represented well by the Curtis-Pennington part (\ref{CP}) of our vertex ansatz. 
This will play an important part in our discussion of the analytical properties 
of the quark propagator at the end of section \ref{complexresults}.

In the quark-DSE the combination of the ghost dressing functions from the 
non-Abelian part of the vertex and the dressing function from the gluon
propagator can be recombined to form the strong running coupling 
in a $\widetilde{MOM}$-scheme, \emph{i.e.} defined from the ghost-gluon vertex:
\begin{equation}
\alpha(k^2) = \frac{g^2}{4\pi} G^2(k^2) Z(k^2)
\,.
\end{equation}
The Dyson-Schwinger equation for the quark propagator then reads
\begin{eqnarray}
  S^{-1}(p) &=& Z_{2}S^{-1}_{0}(p)
  + C_{F}Z_{2}\intq\,  \gamma_{\mu}S(q)\Gamma_{\nu}^{Abel}(q,k)
   P_{\mu \nu} \frac{\alpha(k^2)}{k^2}  \,. \label{DSE2}
\end{eqnarray}
This equation, first developed in \cite{Fischer:2003rp}, is quite similar to
the integral equation Gribov derived his differential equation from. The
quark-gluon interaction is basically given by the strong running coupling
and the dressed vertex is chosen such that it satisfies the Abelian
version of the Slavnov-Taylor identity. Note, however, that the present
approximation is more sophisticated compared to the one of Gribov with respect
to
two points. First, the coupling under the integral is momentum dependent,
whereas Gribov approximated even further by replacing 
$\alpha(k^2) \rightarrow \alpha(0)$.
As a consequence we find the correct leading order anomalous dimensions for
the quark dressing functions in the UV.
Second, the Abelian part of the vertex
nevertheless
satisfies the full WTI as opposed to Gribov's version which satisfied only
the differential form of the WTI valid for zero gluon momentum. For these reasons
we believe that the approximation 
%(\ref{DSE1})
(\ref{DSE2})
is more accurate than the version
of Gribov. 

The explicit expression for $Z(k^2)$ used in this work has been determined in 
ref.~\cite{Alkofer:2003jj} by a fit to numerical solutions of the coupled 
system of DSEs for the ghost and gluon propagators. It is given by
\begin{equation}
Z\left(k^2\right) = \left(\frac{k^2}{k^2+\Lambda^2_{\tt QCD}}\right)^{2\kappa}
\left(\frac{\alpha_{\rm fit}\left(k^2\right)}{\alpha_\mu}\right)^{-\gamma}\;,
\label{gluondress}
\end{equation}
with the gluon momentum $k^2$, the one-loop value 
$\gamma = (-13 N_c + 4 N_f)/(22 N_c - 4 N_f)$ for the anomalous dimension of the 
re-summed gluon propagator and $\alpha_\mu=0.2$ at the renormalisation scale 
$\mu^2=170$~GeV$^2$. We use $\Lambda^2_{\rm QCD}=0.5$~GeV$^2$ similar to the scale 
obtained in ref.~\cite{Alkofer:2003jj}. The infrared exponent $\kappa$ has been 
determined analytically in \cite{Zwanziger:2001kw,Lerche:2002ep} and is given by
$\kappa=\left(93-\sqrt{1201}\right)/98\simeq0.595$.  
The running coupling $\alpha(p^2)$ is parameterised such that the numerical results
for Euclidean scales are accurately reproduced \cite{Alkofer:2003jj}:
\begin{equation}
\alpha_{\rm fit}(p^2) = \frac{\alpha_s(0)}{1+p^2/\Lambda^2_{\rm QCD}}
+ \frac{4 \pi}{\beta_0} \frac{p^2}{\Lambda^2_{\rm QCD}+p^2}
 \left(\frac{1}{\ln(p^2/\Lambda^2_{\rm QCD})}
- \frac{1}{p^2/\Lambda_{\rm QCD}^2 -1}\right)\;. 
\label{alpha}
\end{equation}
Here $\beta_0=(11N_c-2N_f)/3$, and $\alpha_S(0)$ is the fixed point in the
infrared, calculated to be $\alpha_S(0)=8.915/N_c$ for our choice of $\kappa$.
Note that such a fixed point has also been found for the couplings from the 
three-gluon and four-gluon vertices \cite{Alkofer:2004it,Kellermann:2008iw}.

The expressions (\ref{gluondress}) and (\ref{alpha}) represent solutions of the Yang-Mills
part of QCD with important properties. First, note the analytic structure of the
gluon dressing function (\ref{gluondress}) produces a cut along the entire timelike 
$k^2$-axis representing the possibility of the gluon to decay into ghost-antighost
pairs and also into gluons. However, these particles are not physical and need to
be confined. For the gluon dressing function this is reflected in its spectral properties
which have been determined in \cite{Alkofer:2003jj}: the gluon has a positive spectral
function for scales below approximately one fermi, whereas it is negative for larger
scales. As a result the gluon appears to be a free particle in perturbation theory
whereas it cannot propagate freely at larger scales. 

%%%%
%% CF and RW: slightly changed:
The resulting running coupling (\ref{alpha}) has an analytic structure similar
to the one anticipated from analytic perturbation theory \cite{Shirkov:2006gv}.
In addition, it displays an infrared fixed point. Thus in contrast to the setup 
of Gribov, where this infrared fixed point behaviour had to be assumed, we are in 
a position to use an explicitly calculated coupling with the same property. 
Note, however, that this is only possible due to our choice for the quark-gluon 
vertices (\ref{CP}) and (\ref{1BC}). As mentioned above the explicit solution for
this vertex given in \cite{Alkofer:2006gz} is slightly less singular than our ansatz. 
Using the model of \cite{Alkofer:2008et}, which reproduces this behaviour, we have 
checked that this difference has no qualitative impact on most of our results with 
the exception of those reported in section \ref{dsol}, where we will comment further. 
%%%%

\subsection{Pion back-coupling\label{pionsec}}

\begin{figure}[t]
\centerline{\epsfig{file=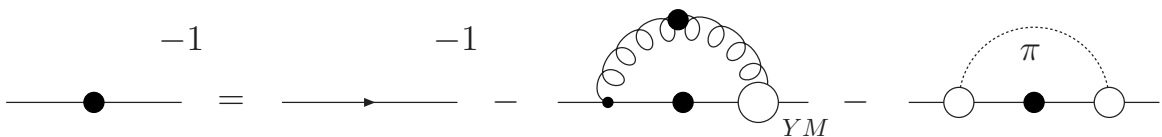,width=16cm}}
\caption{The approximated Schwinger-Dyson equation for the 
quark propagator with effective one-gluon exchange and 
one-pion exchange. 
\label{fig:quarkdse2}}
\end{figure}

As stated in the introduction, Gribov argued that the effects of the
back-reaction by the Goldstone bosons on the quarks should be crucial to 
generate colour-confinement \cite{Gribov:1999ui}. To this end he
%
%assumed 
determined
a form of the pion back-reaction that couples the pion directly to the 
quark. This can be displayed diagrammatically as 
done in fig.~\ref{fig:quarkdse2}.
A similar expression for the pion back-reaction has been derived in 
ref.~\cite{Fischer:2007ze}. Here we develop
a modified approximation scheme leading to a slightly different interaction.

\begin{figure}[t]
\centerline{\epsfig{file=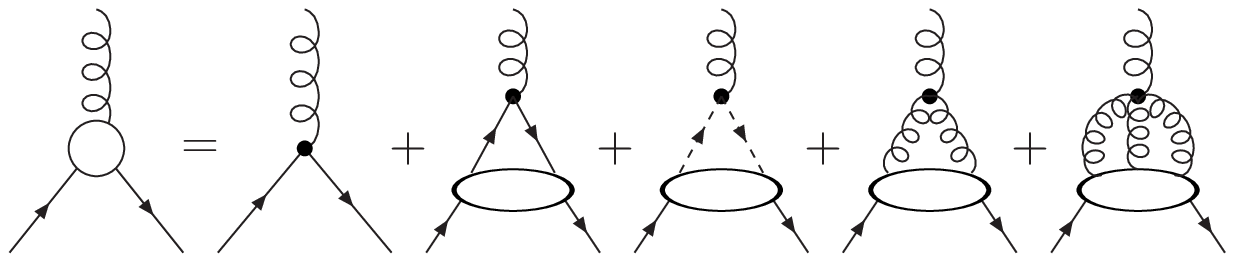,width=15cm}}
\vspace*{5mm}
\centerline{\epsfig{file=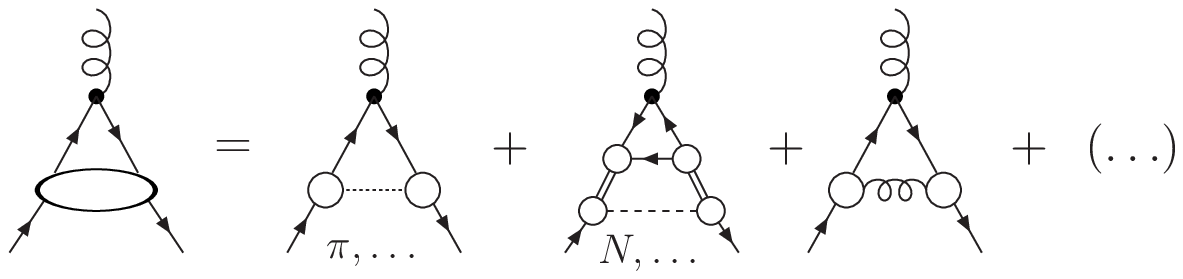,width=13cm}}
\caption{The full, untruncated Dyson-Schwinger equation for the quark-gluon
vertex \cite{Marciano:1977su} is shown diagrammatically in the first line.
The second line describes the first terms of an expansion in terms of hadronic 
and non-hadronic contributions to the quark-antiquark scattering kernel. In both 
equations, all internal propagators are fully dressed. Internal dashed lines with 
arrows correspond to ghost propagators, curly lines to gluons and full lines to 
quark propagators. All internal propagators are fully dressed.
In the second equation, the dotted line describes mesons, the 
dashed line baryons and the double lines correspond to diquarks. }
\label{fig:Vertexdse}
\end{figure}
Consider the Dyson-Schwinger equation of the fully
dressed quark-gluon vertex, given in the first line of fig.~\ref{fig:Vertexdse}. 
For very small momenta, a self-consistent solution 
to this equation has been given in ref.~\cite{Alkofer:2006gz}. Here we are 
primarily interested in the mid-momentum behavior of the vertex and in 
particular in hadronic contributions. To lowest order in a skeleton expansion 
such contributions can only occur in the diagram with the bare quark-gluon 
vertex at the external gluon line. In the second line of 
fig.~\ref{fig:Vertexdse} we expand the quark-antiquark scattering amplitude of this 
diagram in terms of resonance contributions to the kernel and one-particle 
irreducible Green's functions. Amongst other terms discussed in \cite{Fischer:2007ze}
one finds one-meson exchange between the quark and anti-quark lines. Of all
the hadronic contributions this term should be dominant, since diagrams involving heavier
mesons and baryons are suppressed by factors of 
$\Lambda_{QCD}^2/m_H^2$ with $H \in \{K,\rho,N,...\}$.
 
\begin{figure}[t]
\centerline{\epsfig{file=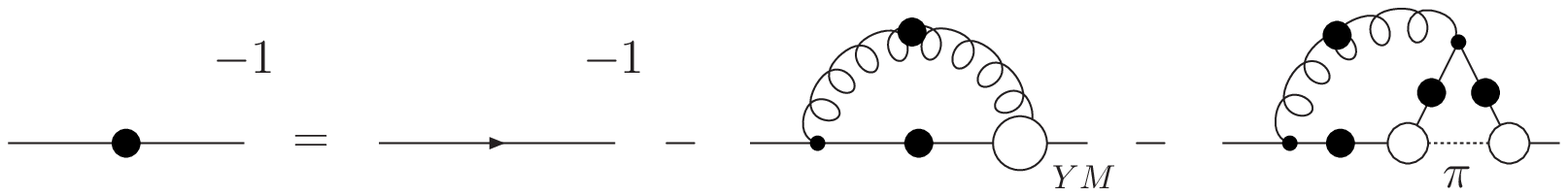,width=15cm}}
\vspace*{5mm}
\centerline{\epsfig{file=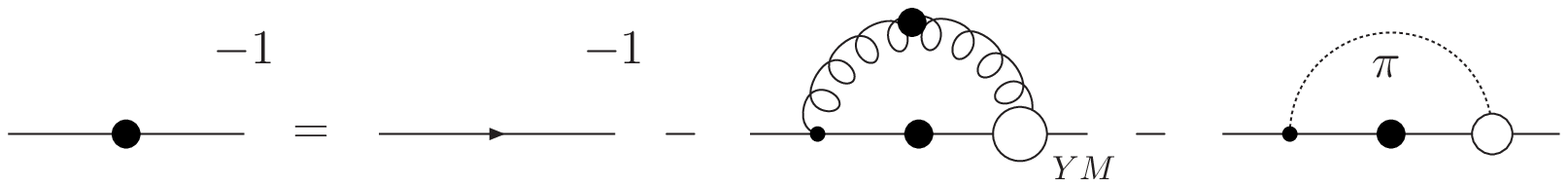,width=15cm}}
\caption{The Schwinger-Dyson equation for the quark propagator
with the quark-gluon vertex from Fig.~\ref{fig:Vertexdse} (upper panel)
and further approximated (lower panel).
\label{fig:quarkdse}}
\end{figure}

Plugging the resulting approximation for the quark-gluon vertex into the 
quark-DSE one arrives at the diagrammatic expression shown in the upper
panel of fig.~\ref{fig:quarkdse}. Here the part denoted by the subscript
`YM' denotes contributions of a purely gluonic nature. This Yang-Mills part
of the interaction has been specified in the previous subsection.
The pion-part is quite
complicated, since it involves not only two-loop integrals but also the
full pion Bethe-Salpeter vertex which needs to be determined from a
Bethe-Salpeter equation.

We will further simplify this expression by noting that one of the loops
involves two bare quark-gluon vertices and a dressed gluon propagator. The
latter one is suppressed at large momenta and at most constant if not
vanishing in the infrared (see \emph{e.g.}
\cite{Lerche:2002ep,Fischer:2002hna} and references therein).
In our earlier work~\cite{Fischer:2007ze} we have approximated this loop by
the Bethe-Salpeter vertex, which would be justified if the full
quark-gluon vertex is almost bare.
However this leads to an over-estimation of the back-reaction and we therefore
only assume it to be proportional to $Z_2 \gamma_5\tau^i$ here.
Indeed, a good agreement with lattice QCD results for the quark propagator and also
meson phenomenology~\cite{mesonpaper} is obtained by setting the loop to be equal
to $Z_2 \gamma_5\tau^i$.
We have checked that the qualitative conclusions drawn in this paper do not
change when employing the truncation used in ref.~\cite{Fischer:2007ze}, which
is actually more similar to Gribov's approach.
One then arrives at the approximated quark-DSE displayed in the lower panel of
fig.~\ref{fig:quarkdse}.

Both, gluon and pion exchange are now given by a one-loop diagram with one
dressed and one bare vertex, respectively. Compared to the previous work
of ref.~\cite{Fischer:2007ze}, fig.~\ref{fig:quarkdse2}, one of the dressed 
pion-quark vertices has disappeared. As an effect, the pion back-reaction
onto the quark is somewhat reduced. This is in line with the results of
\cite{Fischer:2007ze}, where it has been found that the interaction of
fig.~\ref{fig:quarkdse2} leads to far too strong back-reaction effects which
finally resulted in a dramatically small pion decay constant.
Furthermore, our new approximation removes potential
problems with double counting vertex contributions, which are generically
present in DSEs with all vertices dressed. For the purpose of the present paper
we will use the approximation of fig.~\ref{fig:quarkdse}, though we also
performed calculations with both pion-quark vertices dressed.
 
In principle, the pion in the loop couples to the quark line with its full 
Bethe-Salpeter vertex function at the dressed vertex. In general
this function can be decomposed into four different tensor structures
\begin{equation}
\Gamma^i_{\pi}(p,P) = \tau^i \gamma_5 \left\{ E_\pi(p,P) - i \Pslash\, F_\pi(p,P) 
                    - i \pslash \,p \cdot P\, G_\pi(p,P) 
                    - [\Pslash,\pslash] \,H_\pi(p,P) \right\}\,, \label{pion}
\end{equation}
where $\tau^i$ denotes the flavour structure of the vertex, $p$ is the relative
and $P$ the total momentum of the bound state. This pion bound state is the pole
contribution of the full pseudoscalar vertex function. In the chiral limit an exact
solution for the functions $E_\pi,F_\pi,G_\pi,H_\pi$ in terms of the quark self 
energies and regular parts of the isovector axial-vector vertex has been given 
in \cite{Maris:1997hd}. For the leading part $E_\pi$ of the vertex the solution
in the chiral limit depends on the scalar part $B(p^2)$ of the quark propagator
and the pion decay constant $f_\pi$ and is given by $E_\pi(p,P) = B(p^2)/f_\pi$
such that the pion vertex in this approximation reads
\begin{equation}
\Gamma^i_{\pi}(p,P) = \tau^i \gamma_5 \frac{B(p^2)}{f_\pi} \label{piapprox}  
\,.
\end{equation}
For the given truncation scheme two of us have checked explicitly that this
expression is also a very good approximation to the full amplitude $E_\pi$
for a pion with realistic mass, see the appendix of ref.~\cite{mesonpaper}. 
We therefore use (\ref{piapprox}) for the pion vertex in the second diagram
of fig.~\ref{fig:quarkdse}. The resulting complete quark-DSE then reads
\begin{eqnarray}
  S^{-1}(p) &=& Z_{2}S^{-1}_{0}(p)
  + C_{F}Z_{2}\intq\,  \gamma_{\mu}S(q)\Gamma_{\nu}^{Abel}(q,k)
   P_{\mu \nu} \frac{\alpha(k^2)}{k^2} \nonumber\\ 
  &&\hspace*{2cm} - 3 Z_2 \intq\,\gamma_5 S(q) \gamma_5
  \frac{B((p+q)^2/4)}{f_\pi} \frac{1}{k^2+m_\pi^2}  
  \,,
   \label{DSE3}
\end{eqnarray}
with $k=q-p$ and a factor of three in the pion interaction part due to the
flavour factors. 

The pion part of the quark-DSE can now be compared with the one Gribov suggested 
in ref.~\cite{Gribov:1999ui}. There are two differences. The first one concerns the 
appearance of two dressed pion-quark vertices in the back-reaction diagram considered
by Gribov, whereas 
our approximation only includes one dressed vertex for reasons discussed above.
The second concerns the form of the pion Bethe-Salpeter amplitude. Here Gribov 
considered the form
\begin{equation}
\Gamma^i_{\pi}(p,P) = \tau^i \left\{\gamma_5, S^{-1}(p)\right\}/f_\pi
\,,
\label{gribpi}
\end{equation}
which involves further structure in the pion amplitude besides the leading 
$\gamma_5$-part considered in eq.~(\ref{piapprox}). Unfortunately these
additional 
terms are proportional to the quark dressing function $A(p^2)$ and therefore
have the wrong asymptotics at large momenta as compared to actual solutions
of the pion Bethe-Salpeter equation. 

Finally we need to specify the values of the pion mass $m_\pi$ and decay constant 
$f_\pi$ for the pion propagator in the DSEs. In the chiral limit the pion is
a Goldstone boson and $m_\pi=0$ MeV. We use this value together with $f_\pi=90$ MeV.
Away from the chiral limit we use the physical values $m_\pi=138$ MeV and
$f_\pi=93$ MeV for simplicity. We explicitly checked that the qualitative features of
all our results do not depend on variations of these numbers. The quantitative effects
are very small.

\subsection{Renormalisation procedure}

Before we solve equation (\ref{DSE3}) we have to specify our renormalisation
procedure. Upon multiplying (\ref{DSE3}) with $1_{4x4}$ and ${\pslash}$
respectively 
and taking the Dirac trace, one projects the equation onto the self-energies
$B(p^2)$ and $A(p^2)$ contained in the fully dressed quark propagator
$S^{-1}(p)=i {\pslash} A(p^2) + B(p^2)$. Schematically one obtains
\begin{eqnarray}
B(p^2;\mu^2) &=& Z_2(\mu^2) \,m + Z_2(\mu^2)\, \Pi_B(p^2;\mu^2)\,, \label{B}\\
A(p^2;\mu^2) &=& Z_2(\mu^2)   + Z_2(\mu^2) \,\Pi_A(p^2;\mu^2) \,,
\end{eqnarray}
where we have made the dependence on the renormalisation point $\mu^2$
explicit. 
The renormalisation factor $Z_2$ is then determined by
evaluating the second equation at the renormalisation point, \emph{i.e.}
\begin{eqnarray}
Z_2(\mu^2) &=& \frac{A(\mu^2;\mu^2)}{1 +  \Pi_A(\mu^2;\mu^2)} \,,
\end{eqnarray}
with the renormalisation condition $A(\mu^2;\mu^2)=1$. In a numerical
iterative procedure this is always the first step at every
iteration step.
Furthermore, away from the chiral limit $m=0$ one can 
eliminate the renormalisation point independent mass parameter 
$m=m(\mu^2) Z_m(\mu^2)$ by subtracting (\ref{B})
at the renormalisation point. This results in
\begin{equation}
B(p^2;\mu^2) = B(\mu^2;\mu^2) + Z_2(\mu^2) \,
\left(\Pi_B(p^2;\mu^2) - \Pi_B(\mu^2;\mu^2)\right)
\,,
\end{equation}
with the input mass 
$B(\mu^2;\mu^2) = M(\mu^2;\mu^2)\,A(\mu^2;\mu^2)= M(\mu^2;\mu^2)$ 
at the renormalisation point $\mu^2$.

\subsection{Quark propagator in the complex plane \label{complex}}

The behaviour of the quark propagator in the complex momentum plane
and the associated analytic structure of the propagator can be investigated
in two ways. One possibility is to read off the analytic structure 
from the corresponding Schwinger function
\begin{equation}
\sigma_{S,V}(t)\; =\; \int d^3x \int \frac{d^4p}{(2\pi)^4} 
\,e^{i p \cdot x} \,\sigma_{S,V}(p^2)
\,,
\end{equation}
where $\sigma_{S,V}$ are the scalar and the vector parts, respectively, 
of the dressed quark propagator, \emph{i.e.} $\sigma_{S}(p^2)=B(p^2)/\left(p^2
A^2(p^2) + B^2(p^2)\right)$ and
$\sigma_{V}(p^2)=A(p^2)/\left(p^2 A^2(p^2) + B^2(p^2)\right)$. This method has a long history, see 
\cite{Burden:1991gd,Oehme:1994pv,Burden:1997ja,Alkofer:2003jj} and references therein. 
According to the Osterwalder-Schrader axioms of 
Euclidean field theory \cite{Osterwalder:1973dx}, the function $\sigma_{S,V}(t)$ 
has to be positive to allow for asymptotic quark states in the physical sector 
of the state space of QCD. Conversely, positivity violations in the Schwinger function
show that the corresponding asymptotic states (if present) belong to the
unphysical part of the state space. Thus positivity violations constitute a 
sufficient condition for confinement. Moreover, by fitting $\sigma_{S,V}(t)$ with appropriate
forms one obtains information on the dominant (\emph{i.e.} closest to the origin) 
non-analyticity of the quark propagator in the complex plane. In this work we use 
the form 
\begin{equation}
\sigma_{S,V}(t)\; =\; \, b_0 \,e^{-b_1 t} \,\cos(b_2 t+b_3)\, , \label{cc}
\end{equation}
which corresponds to a pair of complex conjugate poles of the propagator 
in the timelike momentum plane located at $m_{pole} = b_1 \pm i \, b_2$. The
Schwinger function is then oscillating around zero with periodicity $b_2$.
If $b_2=0$
one obtains an exponentially damped Schwinger function corresponding
to a pole on the real negative momentum axis at $m_{pole} = b_1$ 
(see ref.~\cite{Alkofer:2003jj} for more details). Thus by calculating the Schwinger
function once with and once without the pion back-reaction we have a reliable tool to
assess possible changes in the analytical structure of the quark propagator. This 
allows us to test Gribov's conjecture. 

These findings can be further corroborated by a direct calculation of the quark
propagator in the complex plane. Technically, however, there is a caveat. Consider
\emph{e.g.} the explicit form of the DSE for the scalar self-energy $B(p^2)$ using the
1BC-vertex defined in eq.~(\ref{1BC}):
\begin{eqnarray}
B(p^2) = Z_2 \, m &+& \frac{C_f Z_2}{4 \pi^3} \int d^4q \,
\frac{\alpha(k^2)}{k^2} \,\frac{3\, B(q^2)}{q^2 A^2(q^2)+B^2(q^2))} 
\frac{A(p^2)+A(q^2)}{2} \nonumber\\
&-& Z_2 \frac{3}{(2\pi)^4} \int d^4q \,
\frac{1}{k^2+m_\pi^2} \,\frac{B(q^2)}{q^2 A^2(q^2)+B^2(q^2))} 
\frac{B((p+q)^2/4)}{f_\pi}\,. \label{eq1}
\end{eqnarray}
A similar equation holds for $A(p^2)$.
Solving this equation directly at complex momenta $p^2$ entails a complex argument
$k^2=(p-q)^2$ of the running coupling $\alpha(k^2)$. This coupling, however, has
its own analytic structure, given by eq.~(\ref{alpha}). Although (\ref{alpha})
represents a reasonably motivated and justified guess for the analytic structure of
the coupling we would rather avoid relying upon it. Fortunately it turns
out that this can be easily accomplished. To this end we shift the loop momentum $q^2$ in 
the quark-DSE such
that the complex argument $k^2$ does not appear in the gluon part of the loop but 
instead runs through the internal quark part of the loop. For the shifted equation we obtain
\begin{eqnarray}
B(p^2) = Z_2 \, m &+& \frac{C_f Z_2}{4 \pi^3} \int d^4q \,
\frac{\alpha(q^2)}{q^2} \,\frac{3\, B(k^2)}{k^2 A^2(k^2)+B^2(k^2))} 
\frac{A(p^2)+A(k^2)}{2}\nonumber\\
&-& Z_2 \frac{3}{(2\pi)^4} \int d^4q \,
\frac{1}{q^2+m_\pi^2} \,\frac{B(k^2)}{k^2 A^2(k^2)+B^2(k^2))} 
\frac{B((p+k)^2/4)}{f_\pi}\,, \label{eq2}
\end{eqnarray}
and a similarly shifted equation for $A(p^2)$. The running coupling is now evaluated
for real momenta $q^2$ only, whereas the quark propagator is determined self-consistently
for complex momenta $p^2$ and $k^2$.

The only caveat in this procedure is the interference with the regularisation
procedure.
Obviously, if we employ a translationary invariant regularisation scheme such as dimensional
regularisation such a shift would be harmless. In our
numerical procedure, however, we use a hard cut-off scheme. Then it turns out, that
the unshifted and shifted coupled system of quark-DSEs for $A$ and $B$  
precisely give the same results, if $Z_2$ is kept fixed while shifting. In practise
we calculate $Z_2(\mu^2)$ for a given renormalisation point $\mu^2$ from the 
fully converged unshifted equations and plug its value into the shifted equations 
as an input parameter. Both results then agree to numerical accuracy on the real axis.
We then use the shifted DSEs to solve for $B(p^2)$ and $A(p^2)$ in the complex
plane. Here we developed a new numerical algorithm, which is described in detail in
appendix \ref{complexplane}.

\section{Numerical Results \label{results}}

\subsection{Numerical results on the real axis and the Schwinger function \label{complexresults}}

\begin{figure}[b]
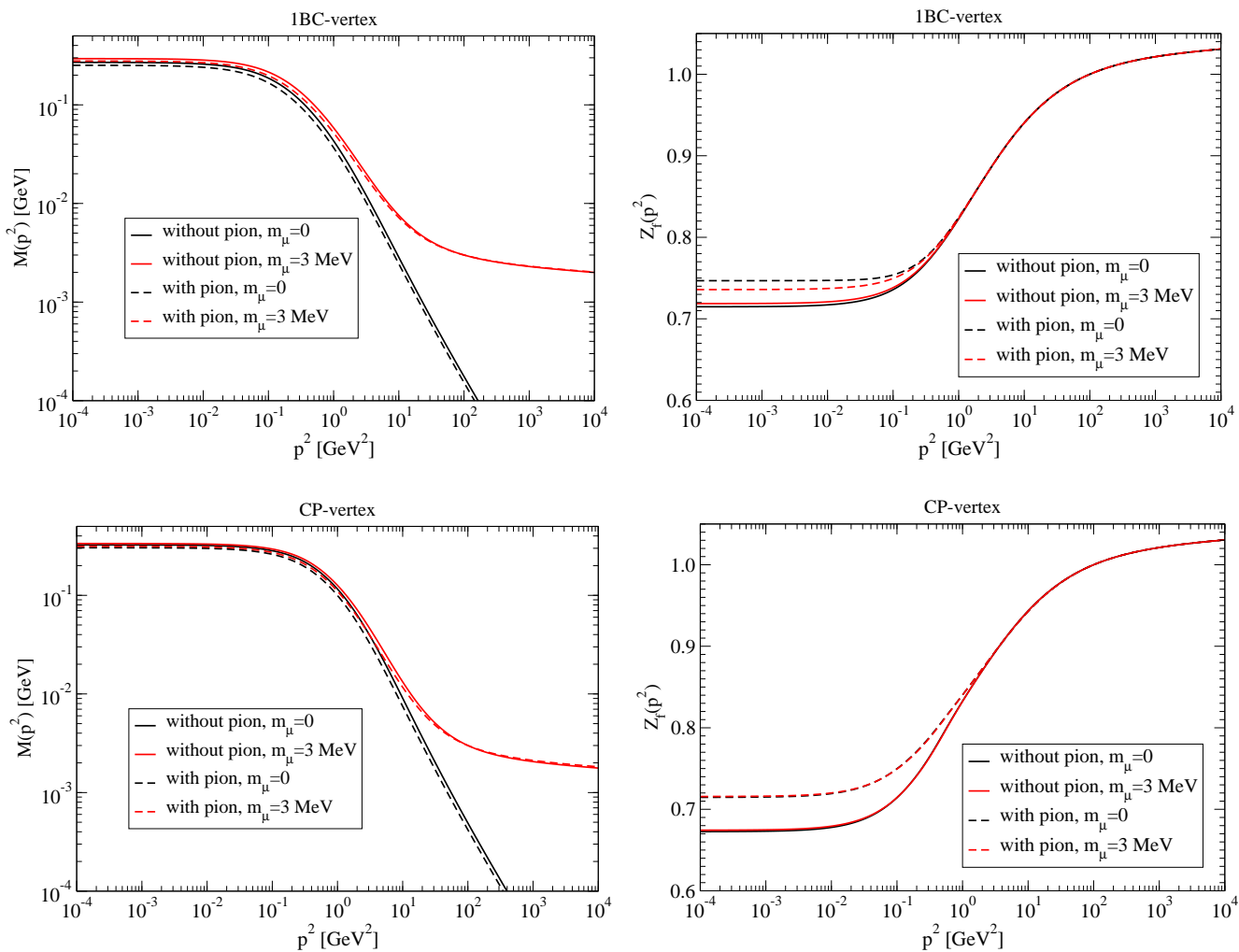

  \vspace{5mm} 
   \centerline{\epsfig{file=M.eps,width=86mm}
    \hspace{1mm}
    \epsfig{file=Zf.eps,width=86mm} }
\vspace*{5mm}
   \centerline{
    \epsfig{file=M.cp.eps,width=86mm}
    \hspace{1mm}
    \epsfig{file=Zf.cp.eps,width=86mm} }
\vspace*{3mm}
\caption{\label{quark}
  The mass function $M(p^2)=B(p^2)/A(p^2)$ of the quark and the wave 
  function $Z_f(p^2)$ with 1BC-vertex (upper panel) and with 
  CP-vertex (lower panel). The explicit mass $m_\mu$ is taken at $\mu=10$ GeV.}
\end{figure}

We first compare the quark mass function $M(p^2)=B(p^2)/A(p^2)$ and the wave
function $Z_f(p^2)=1/A(p^2)$ with and without the pion back-coupling for
real momenta $p^2$. Our numerical results are shown in fig.~\ref{quark}.
In the upper panel we compare results for the simpler 1BC-vertex (\ref{1BC}),
whereas in the lower panel results for the full Curtis-Pennington construction
(\ref{CP}) are shown. In both cases we compare the solutions in the chiral
limit and results
with a small quark mass $m(10 \,\mbox{GeV})= 3 \,\mbox{MeV}$, which roughly
corresponds to an up quark with 
$m_{\overline{MS}}(2 \,\mbox{GeV})= 4 \,\mbox{MeV}$. For both vertex constructions
the results are qualitatively similar. Including the pion back-reaction into the
quark-DSE reduces the amount of dynamical chiral symmetry breaking to some extent.
This reduction is larger in the chiral limit. Explicit values for the quark
mass function at zero momentum are given in table \ref{table} and agree with
this observation.

\begin{table}[t]
\begin{tabular}{|c||c|c|c|}
\hline
          & \hspace*{2mm} $M(0)$ [MeV]           \hspace*{2mm} 
          & \hspace*{2mm} $\left(-\langle \bar{\Psi}\Psi \rangle_{2 \,\mbox{\small GeV}}^{\overline{MS}}\right)^{1/3}$  [MeV] \hspace*{2mm} 
	  & \hspace*{1mm} location of poles $m_{pole}$ [MeV] \hspace*{1mm} 
	                                \rule[-3mm]{0mm}{7mm}\\ \hline\hline
1BC, m=0 MeV, wo $\pi$    & 269  & 216 & 295 (5) $\pm i$ 176 (10) \rule[-3mm]{0mm}{7mm} \\\hline
1BC, m=3 MeV, wo $\pi$    & 294  & -   & 322 (5) $\pm i$ 191 (10) \rule[-3mm]{0mm}{7mm} \\\hline
1BC, m=0 MeV, with $\pi$  & 252  & 208 & 279 (5) $\pm i$ 160 (10) \rule[-3mm]{0mm}{7mm} \\\hline
1BC, m=3 MeV, with $\pi$  & 278  & -   & 304 (5) $\pm i$ 180 (10) \rule[-3mm]{0mm}{7mm} \\\hline\hline
 CP, m=0 MeV, wo $\pi$    & 322  & 289 & 513 (10) $\pm i$   0 (10) \rule[-3mm]{0mm}{7mm} \\\hline
 CP, m=3 MeV, wo $\pi$    & 331  & -   & 530 (10) $\pm i$   0 (10) \rule[-3mm]{0mm}{7mm} \\\hline
 CP, m=0 MeV, with $\pi$  & 299  & 276 & 478 (10) $\pm i$   0 (10) \rule[-3mm]{0mm}{7mm} \\\hline
 CP, m=3 MeV, with $\pi$  & 309  & -   & 493 (10) $\pm i$   0 (10) \rule[-3mm]{0mm}{7mm} \\\hline
\end{tabular}
\vspace{3mm}
\caption{Infrared masses $M(0)$, chiral condensate 
$\left(-\langle \bar{\Psi}\Psi 
\rangle_{2 \,\mbox{\small GeV}}^{\overline{MS}}\right)^{1/3}$
and pole location $m_{pole}$ of the resulting quark propagator determined from fits to
the Schwinger function $\sigma(t)$ for the 1BC-vertex (\ref{1BC}) and CP-vertex (\ref{CP})
choice and two different bare quark masses.\label{table}}
\vspace{5mm}
\end{table}
The ultraviolet behaviour of the quark mass function is given by the
analytic solution \cite{Miransky:1985ib}
\begin{equation}
M(p^2)_{asym} =  \frac{2 \pi^2 \gamma_m}{3}
\frac{-\langle \bar{\Psi}\Psi\rangle}
{p^2 \left(\frac{1}{2} \log(p^2/\Lambda^2_{QCD})\right)^{1-\gamma_m}}
+ \overline{m} \left[\omega\log\left(\frac{p^2}{\Lambda^2_{QCD}}\right)\right]^{-\gamma_m}
\label{asym}
\end{equation}
with the anomalous dimension $\gamma_m = \frac{12}{11N_c-2N_f}$. The quantity
$\overline{m}$ is related to the
current quark mass $m$ in the QCD Lagrangian,
whereas $\langle \bar{\Psi}\Psi\rangle$ is the (renormalisation point independent)
chiral condensate. For $m \not= 0$ the 
dominant part of (\ref{asym}) in the far ultraviolet is the second logarithmic
term,
whereas the $1/p^2$-term is important at intermediate momenta. In the chiral
limit this term is the only one present. From the results of fig.~\ref{quark}
we clearly infer that the condensate term, representing dynamical chiral symmetry
breaking, is modified by the pion back-reaction whereas the logarithmic term,
representing the explicit breaking due to $m$, is not. This is in nice agreement
with our expectations. The explicit term should be largely independent of the details
of the strong interaction, whereas the condensate term is not. One can determine the 
values of the chiral condensate either from fitting (\ref{asym}) to the asymptotics
of the quark mass function or by calculating
\begin{eqnarray}
-\langle \bar{\Psi}\Psi\rangle_\mu := Z_2(\mu^2) \, Z_m(\mu^2) \, N_c 
\,\mbox{tr}_D \int
\frac{d^4q}{(2\pi)^4} S(q^2;\mu^2) \,,
\label{ch-cond}
\end{eqnarray}
in the chiral limit (the trace is over Dirac indices). We determined the condensate
at our renormalisation point $\mu=10$ GeV in the MOM-scheme and converted
it to the conventional $\overline{MS}$-result at $\nu=2$ GeV using
\begin{equation}
-\langle \bar{\Psi}\Psi\rangle^{\overline{MS}}_{2 \,\mbox{\small GeV}} = 
-\langle \bar{\Psi}\Psi\rangle^{MOM}_{10 \,\mbox{\small GeV}} 
\left(\frac{\ln(4 \,\mbox{GeV}^2/\Lambda^2_{\overline{MS}})}
{\ln(100 \,\mbox{GeV}^2/\Lambda^2_{MOM})}\right)^{\gamma_m}
\,,
\end{equation}
with $\gamma_m = \frac{12}{11N_c-2N_f}$ and the scales 
$\Lambda_{\overline{MS}} = 225$~MeV and $\Lambda_{MOM} = 500$~MeV.
We then find the values given in table \ref{table} which support our qualitative 
findings from fig.~\ref{quark}. Note that the unquenching effects due to the pion
back-reaction are small, \emph{i.e.} of the order of 10 MeV in the (third root
of the) condensate. This agrees
with previous findings both in the DSE framework \cite{Fischer:2003rp,Fischer:2005en}
and in lattice calculations \cite{McNeile:2005pd}.

\begin{figure}[t]
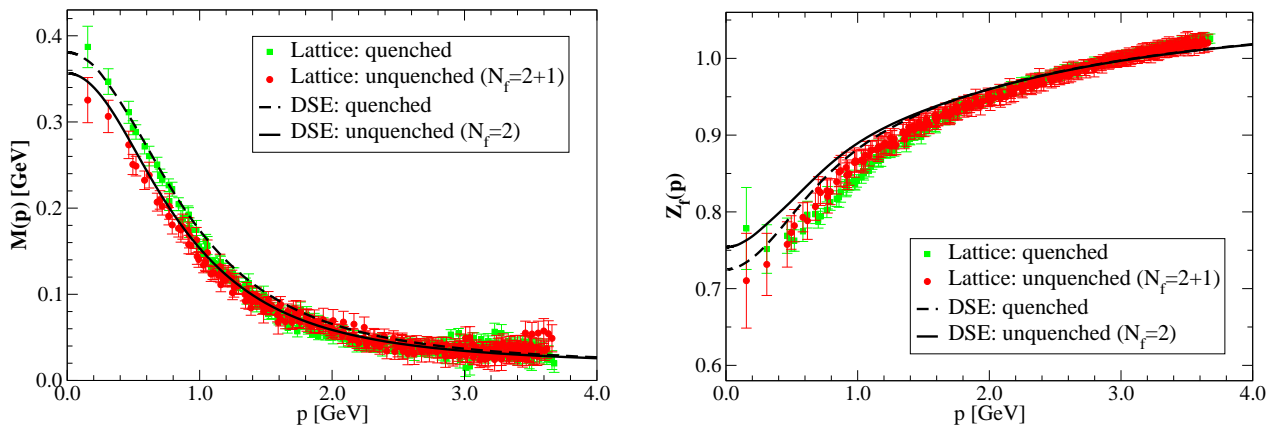

  \vspace{5mm} 
   \centerline{\epsfig{file=latt-M.eps,width=80mm}
    \hspace{5mm}
    \epsfig{file=latt-Z.eps,width=80mm} }
\caption{\label{lattice}
  The mass function $M(p^2)=B(p^2)/A(p^2)$ of the quark and the wave 
  function $Z_f(p^2)$ with CP-vertex compared to the lattice results of 
  Bowman et al. \cite{Bowman:2005vx}. The current quark masses employed on
  the lattice compare to $m(\mu^2)=16$ MeV with $\mu=10$ GeV in our momentum
  subtraction scheme.}
\end{figure}

The effects in the quark mass	
function and the quark wave function are compared with recent lattice results
of Bowman \emph{et al.} \cite{Bowman:2005vx} in fig.~\ref{lattice}. 
The current quark masses employed on the lattice compare to $m(\mu^2)=16$ MeV 
with $\mu=10$ GeV in our momentum subtraction scheme. In the plot our result 
for $Z_f(p^2)$ is renormalised to $\mu=3$ GeV by a finite renormalisation group 
transformation. $M(p^2)$ is a renormalisation group invariant. One finds very good
qualitative and also quantitative agreement of the effects in the quark mass function. 
In the wave function we only find small effects which quantitatively agree with the
effects on the lattice, however with a different sign. A similar difference has 
already been observed in ref.~\cite{Fischer:2005nf} and should be clarified in future
work. Apart from this small deviation we therefore consider
the interaction defined in section \ref{pionsec} to realistically reproduce the
pion back-reaction effects (as opposed to the stronger one considered 
in \cite{Fischer:2007ze})\footnote{From the linear plot one can again clearly infer that 
the pion back-reaction of the quarks is largest in the infrared momentum region. This
is in marked contrast to the findings of previous attempts to quantify the pion
corrections \cite{Ewerz:2006vw,Kumar:2007he}. We attribute this difference to the wrong
asymptotics of the pion wave function (\ref{gribpi}) in Gribov's
equation.}.
We wish to stress that the agreement of our results with the lattice data underlines the
eligibility of our approach to critically assess the influence of Goldstone bosons 
on the analytic structure of the propagator.

\begin{figure}[t]
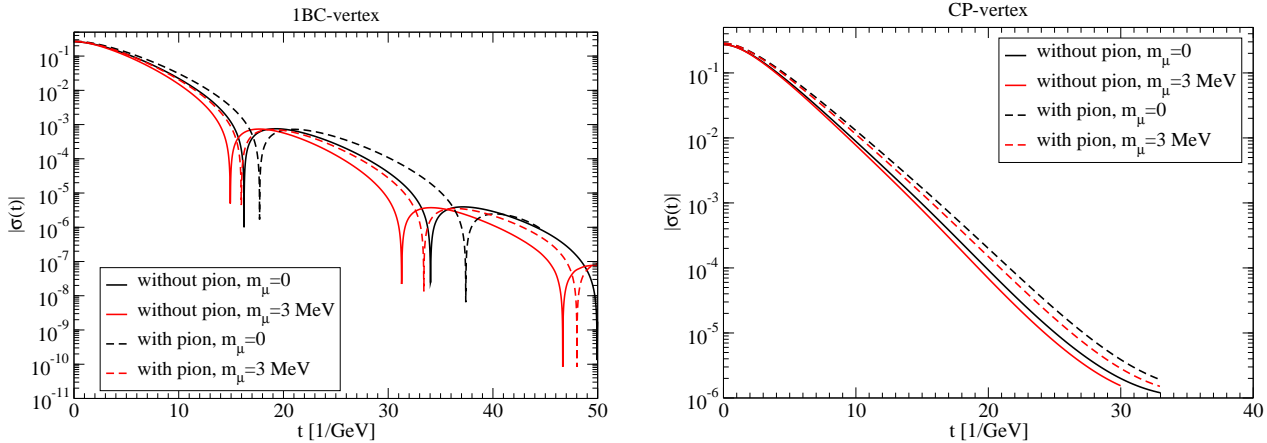

  \vspace{5mm} 
   \centerline{\epsfig{file=FT.eps,width=80mm}
    \hspace{5mm}
    \epsfig{file=FT.cp.eps,width=80mm} }
\caption{\label{FT}
  The absolute value of the Schwinger function $\sigma(t)$ with
  1BC-vertex (left diagram) and with CP-vertex (right diagram).}
\end{figure}

In fig.~\ref{FT} we display the results from the Fourier-transform of the scalar part
of the quark propagator. On a logarithmic scale we show the resulting Schwinger function
$\sigma(t)$ for the two different choices for the quark-gluon vertex once without
and once with the pion back-reaction. It is evident from the results that the pion
back-reaction does not change the analytic structure of the quark propagator. Instead,
it is the form of the quark-gluon vertex that is crucial for the form of the
Schwinger function. If only the first term of the Ball-Chiu solution of the Abelian
WTI is used, eq.~(\ref{1BC}) we obtain an oscillating Schwinger function. In the plot 
this oscillation
is manifest in the vertical spikes. An ansatz representing a pair of complex conjugate
poles in the momentum plane fits such a behaviour nicely and one obtains the pole
locations reported in table \ref{table}. All imaginary parts are clearly significant
and have roughly half the size of the real parts of $m_{pole}$. The inclusion of the
pion back-reaction here diminishes the real parts by about 20 MeV and the imaginary parts
by a somewhat smaller 10 MeV. The situation is totally different when the vertex construction
(\ref{CP}) including the Curtis-Pennington ansatz is used. The relative strengths
of the different tensor structures of the vertex are now such that the Abelian WTI
is satisfied. From the right diagram in fig.~\ref{FT} we see that the resulting Schwinger
function now decays exponentially without any visible oscillations. Such a behaviour is
characteristic for a quark propagator with a singularity on the real axis which may or
may not be accompanied by a cut. The location of this singularity is also reported in
table \ref{table}. Again the inclusion of the pion back-reaction does not change
the qualitative behaviour of the Schwinger function but merely shifts the location
of the singularity by 20-30 MeV to lower values.

This is the central result of this work: the inclusion of the pion back-reaction does not
change the analytical structure of the quark propagator. Instead, as has been discussed
previously in \cite{Alkofer:2003jj}, it is the relative strength of the scalar and vector
terms in the Yang-Mills part of the quark-gluon interaction which is crucial. 
Following ref.~\cite{Alkofer:2003jj} we can interpret this result. Consider for 
a moment the Abelian theory, \emph{i.e.} QED. The non-Abelian part $W^{\neg abel}$ of the 
vertex is then absent and the Abelian WTI (\ref{WTI}) exact. Gauge invariance
then dictates a form of the fermion-photon vertex like the
Curtis-Pennington construction. The resulting physical electron propagator
then is expected to have a singularity at the electron mass accompanied by
a cut due to the accompanying soft photon cloud. This agrees well with our findings.
The quark-gluon vertex in the non-Abelian theory, however, is necessarily
modified compared to the Abelian interaction. This can be seen from its Slavnov-Taylor
identity
\begin{equation}
G^{-1}(k^2) \: k_\mu \: \Gamma_\mu(q,k) = S^{-1}(p) \: H(q,p) - H(q,p) \: S^{-1}(q),
\label{quark-gluon-STI}
\end{equation}
where $G(k^2)$ is the dressing function of the Faddeev-Popov ghosts and $H(q,p)$ 
the ghost-quark scattering kernel. It is currently not clear, though a matter of
current investigations \cite{Alkofer:2006gz,GimenoSegovia:2008sx} whether these
modifications lead to either of the singularity structures displayed in fig.~ 
\ref{FT}. The answer to this question remains important, since oscillations
in the Schwinger function would be a sufficient condition for quark confinement 
as discussed in subsection \ref{complex}.

\subsection{Numerical results for additional solutions of the quark-DSE \label{dsol}}

For sufficiently strong coupling multiple solutions of the quark-DSE exist in
a domain $\mathcal{D} = \left\{ m : 0 \le m \le m_{cr}\right\}$
of the current quark-mass~\cite{Chang:2006bm,Williams:2006vva}.
The appearance of multiple solutions is not surprising and has strong
similarities with hysteresis in ferromagnets.
In his work Gribov emphasized however that his equations for a supercritical
coupling allow for presumably infinite different solutions in the chiral
limit~\cite{Gribov:1992tr}.
This is in contrast to the finding in DSE investigations using the Maris-Tandy
model~\cite{Chang:2006bm,Williams:2006vva}.
Here we explore these multiple solutions
without the inclusion of the pion back-reaction, indicating that the
different behaviour of these multiple solutions is a result of the
truncation scheme employed and the form of the gluon interaction.

All of these multiple solutions are connected to the perturbative running of
quark 
mass at large momenta, differing only in their infrared behaviour.
They can be distinguished by the number of zero crossings that occur in the
quark self-energy $B$, or equivalently in the mass function $M=B/A$.
The energetically preferred or physical solution is strictly positive definite.
Without explicit symmetry breaking, \emph{i.e.} in the chiral limit, the
symmetric
solution without dynamical mass generation has to exist as well.

\begin{figure}[t]
    \centerline{ \epsfig{file=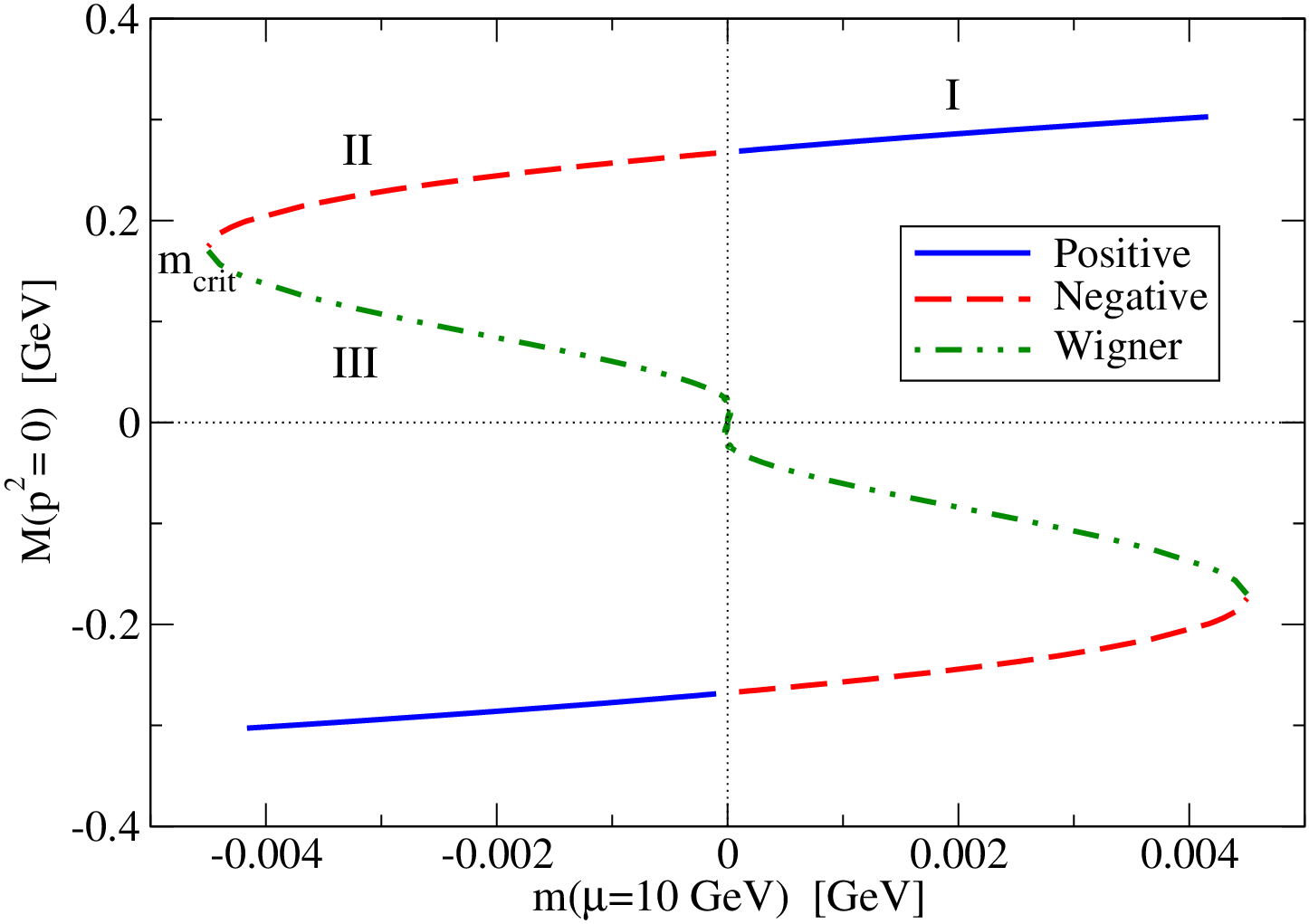,height= 5.5cm} \hspace{.5cm}
		     \epsfig{file=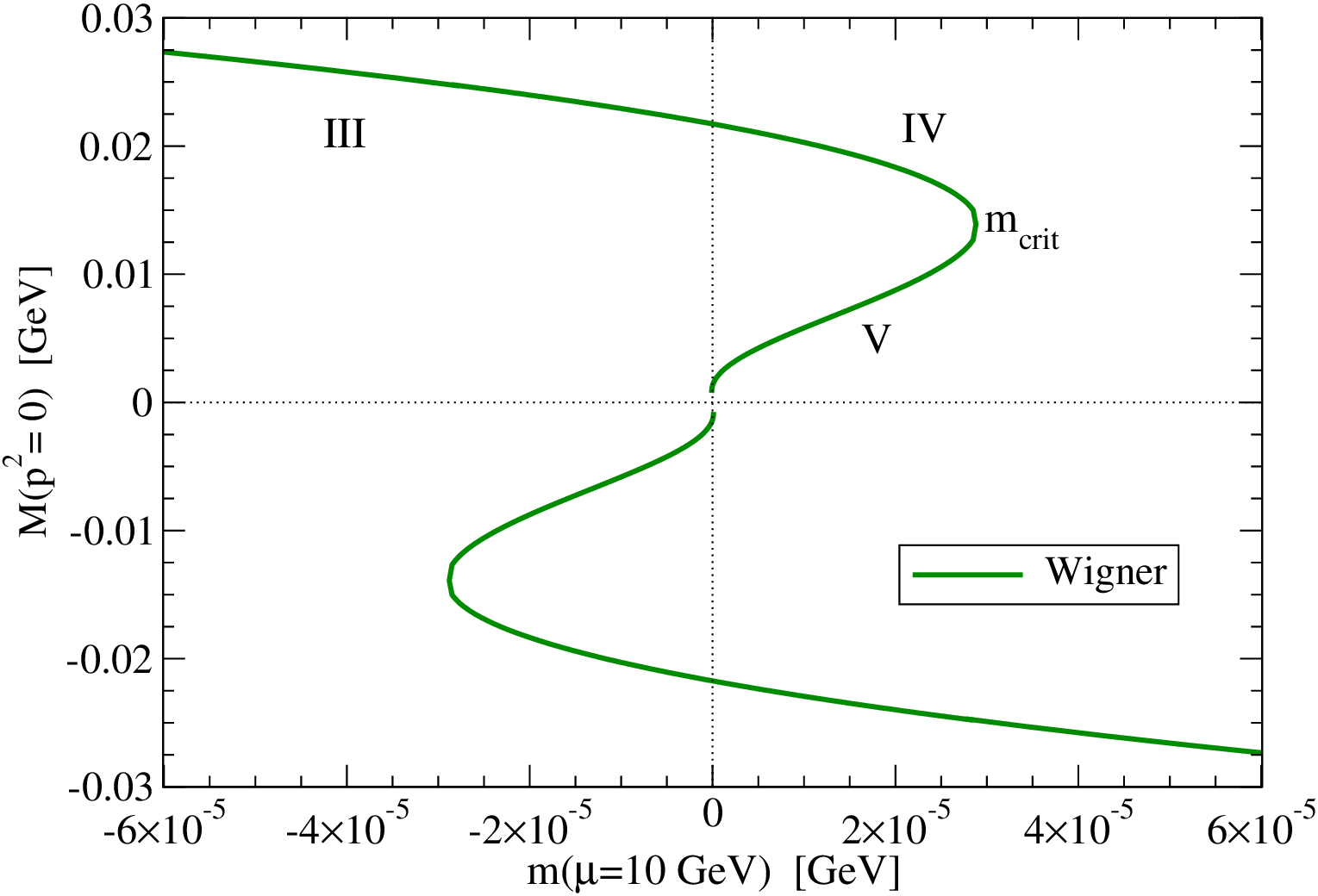,height= 5.5cm}     }
\caption{\label{othersoln}
We show the dynamically generated mass $M(p^2=0)$ for various solutions of the
quark-DSE as a function of the current quark-mass. 
The diagram on the right shows the behaviour of $M(m)$ close to the origin.}
\vspace*{3mm}
    \centerline{ \epsfig{file=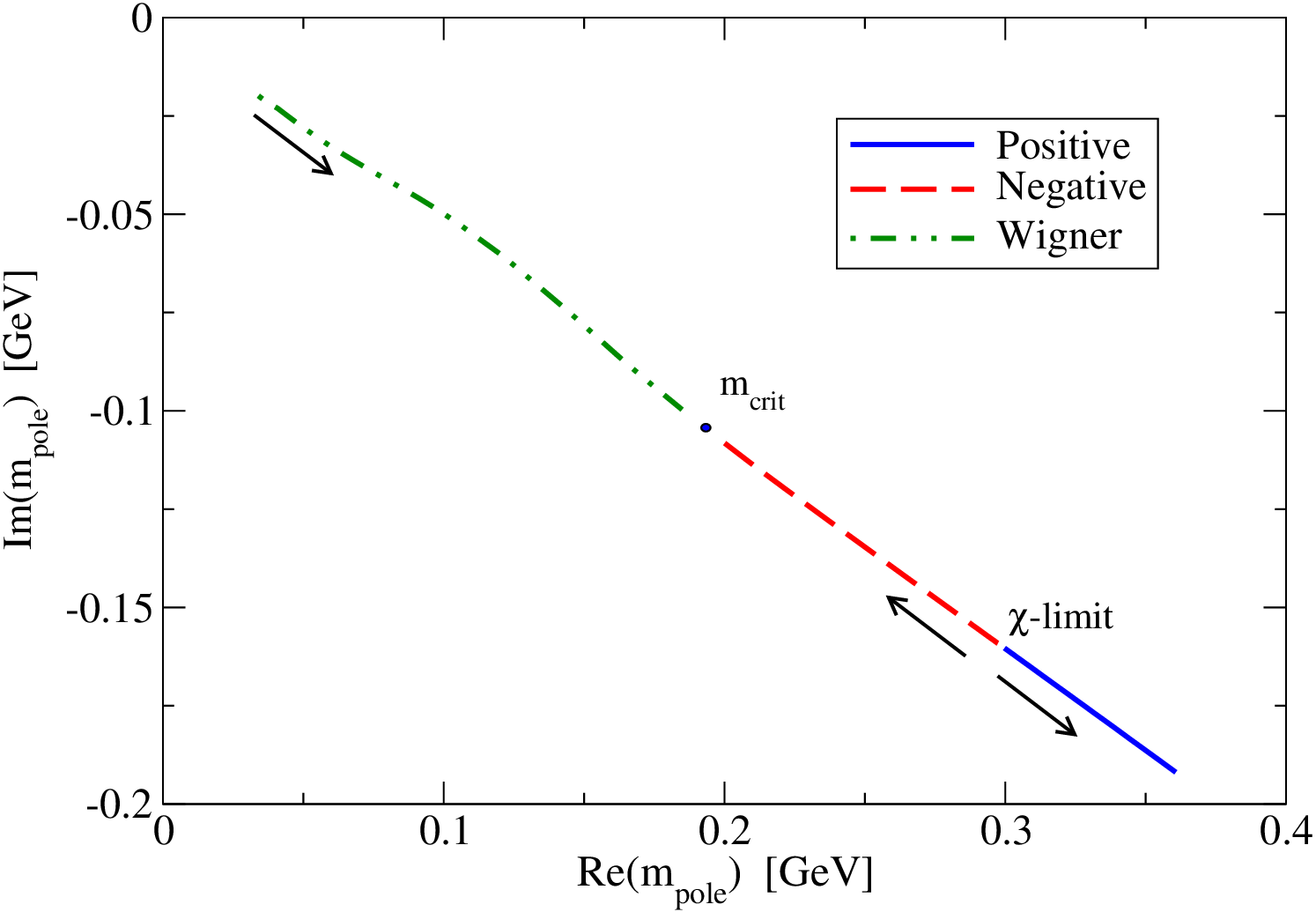,height= 5.5cm} \hspace{.5cm}
		     \epsfig{file=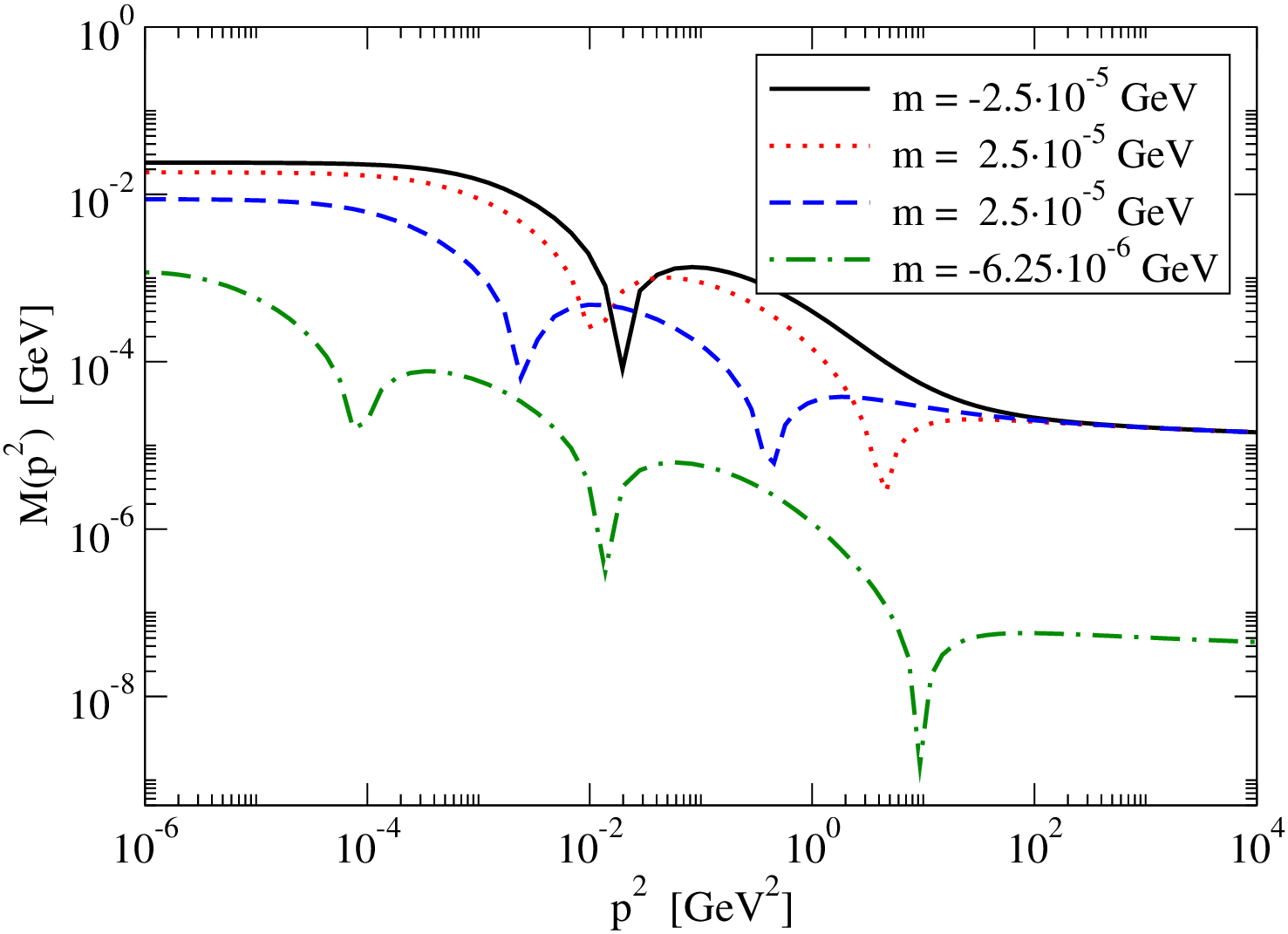,height= 5.5cm}     }
\caption{\label{othersoln2}
In the left diagram we show the real versus imaginary parts of the
complex pole-mass $m_{pole}$ for different solutions of the quark-DSE. Arrows indicate
increasing current quark mass. On the right we present nodal solutions of the mass 
function for the indicated current quark-masses.
The solution with one node corresponds to region III in the right diagram of 
fig.~\ref{othersoln}, the dotted solution corresponds to region IV, the dashed
curve to region V and the dash-dotted to a yet additional solution close to the 
origin of the $M-m$-plane.  
}
\end{figure}

In fig.~\ref{othersoln} and~\ref{othersoln2} we present various properties of
the multiple solutions.
In fig.~\ref{othersoln} we follow the value of the mass function $M(0)$
starting with a positive definite solution beyond $m_{cr}$, denoted `positive'
in fig.~\ref{othersoln} and marked with a `I'.
Decreasing the current quark-mass to zero, \emph{i.e.} going to the chiral limit,
the positive solution is degenerate with a negative solution, whose mass
function has the opposite sign.
This is related to a $\mathbb{Z}_2$ symmetry of the quark-DSE when
simultaneously sending $m\rightarrow -m$ and $M(p^2)\rightarrow -M(p^2)$.
This symmetry is also manifest in fig.~\ref{othersoln}.
Restricting our attention to $M(0)$ positive, we now introduce a
\emph{negative} current quark mass, giving rise to a different pattern of
dynamical mass generation and the so-called `negative' solutions, indicated by
a `II'.

The location of the complex poles, as obtained through the Schwinger function
is shown in the left diagram of fig.~\ref{othersoln2} and reveals that these positive and
negative solutions are smoothly connected. The key difference, however,
is that these negative solutions develop a node in the mass function
as we cross $m=0$ -- that is to say they are no longer positive
semi-definite with the scalar self-energy changing sign for small
($p<10$~GeV) 
momenta. On continuing to increase the magnitude of the negative mass, we
reach a critical point $m_{cr}$ whose model-dependent value here is
approximately $43$~MeV at $\mu=5$~GeV for the 1BC vertex and the
interaction described in section \ref{interaction}. This critical mass merely indicates a
bifurcation where the negative solution is degenerate with the so-called
Wigner solution. At this point, we move again towards the chiral limit,
this time following a different path marked by a `III'.

%%%%
%% CF and RW: slightly changed:
Interestingly, the Wigner solution does not connect immediately to
the trivial solution $M(0)=0$ at $m=0$.
Instead, we find $M(0)>0$ for $m=0$ and therefore a second non-trivial
solution in the chiral limit. 
This behaviour was not observed in previous studies\footnote{We did, in
fact, find two-noded solutions with the Maris-Tandy interaction.
However, these have poles on the negative real-axis and do not smoothly
connect to the positive solution as we find in fig.~\ref{othersoln2}.} of
these solutions~\cite{Chang:2006bm,Williams:2006vva}.
There, extensive use of the Maris-Tandy~\cite{Maris:1999nt} interaction has
been made. Also the soft-divergence model 
of \cite{Alkofer:2008et} constructed to exhibit the infrared properties of
the quark-gluon vertex determined in \cite{Alkofer:2006gz} does not show the
second class of non-trivial Wigner solutions observed in the right diagram of 
fig.~\ref{othersoln}. Both the Maris-Tandy model and the soft-divergence model 
vanish at zero momentum with different powers of momentum squared. The behaviour
of fig.~\ref{othersoln} is only present when the quark-gluon vertex is given 
enough strength leading to either an infrared-fixed point or a singular behaviour
in the effective running coupling, as is the case in our interaction and that of
Gribov's.
%%%%

This second Wigner solution develops a total of two 
zero-crossings in the mass function as we cross $m=0$.
Increasing the mass, we follow curve `IV', which  
again bifurcates into two solutions at some second critical mass
$m\simeq0.029MeV$.
Continuing the procedure we follow path `V' and again cross $m=0$ 
with the now characteristic development of an additional node in the mass function (see
fig.~\ref{othersoln2}).
At this point, we make no further attempt to resolve such solutions
since the critical mass now oscillates around $m=0$ with rapidly decreasing
amplitude. We can only presume that such solutions continue to exist,
with yet more nodes developing in the mass-function as the trivial
solution is approached. We expect the
multitude of these solutions to smoothly connect the location of the 
complex poles to the trivial solution, as indicated in the left diagram of 
fig.~\ref{othersoln2}.  

Once again, we point out that we did not include pion effects here, due
in part to the ambiguity of choosing $M_\pi$ and $f_\pi$ for varying
quark mass. Since Gribov found the same results in his approximation
with the pion back-reaction~\cite{Gribov:1992tr}, it seems likely that the
inclusion of pions does not change this picture quantitatively.

\subsection{Numerical results in the complex plane}
\begin{figure}[t]
\centerline{\epsfig{file=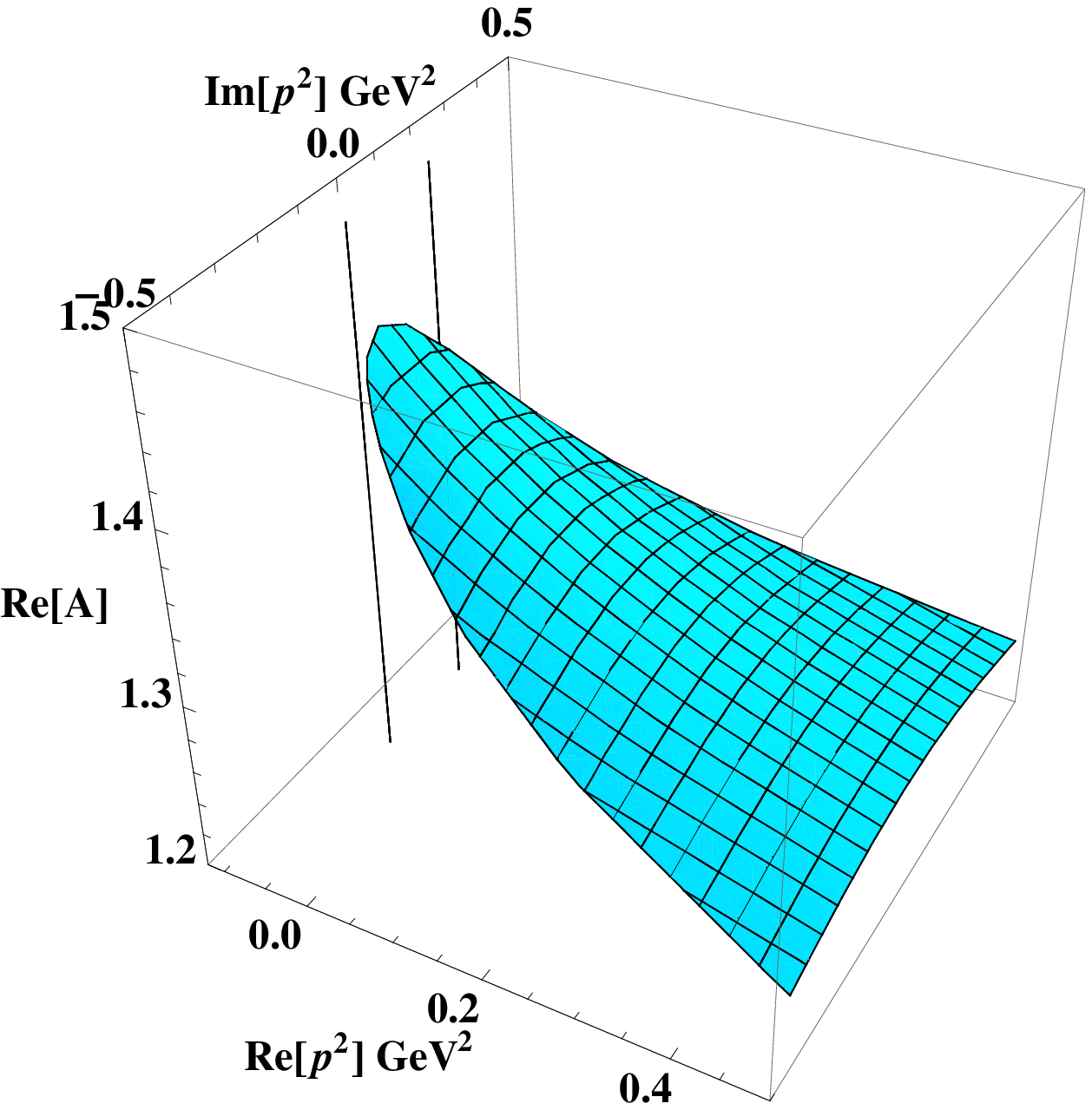,width=7cm}
		\epsfig{file=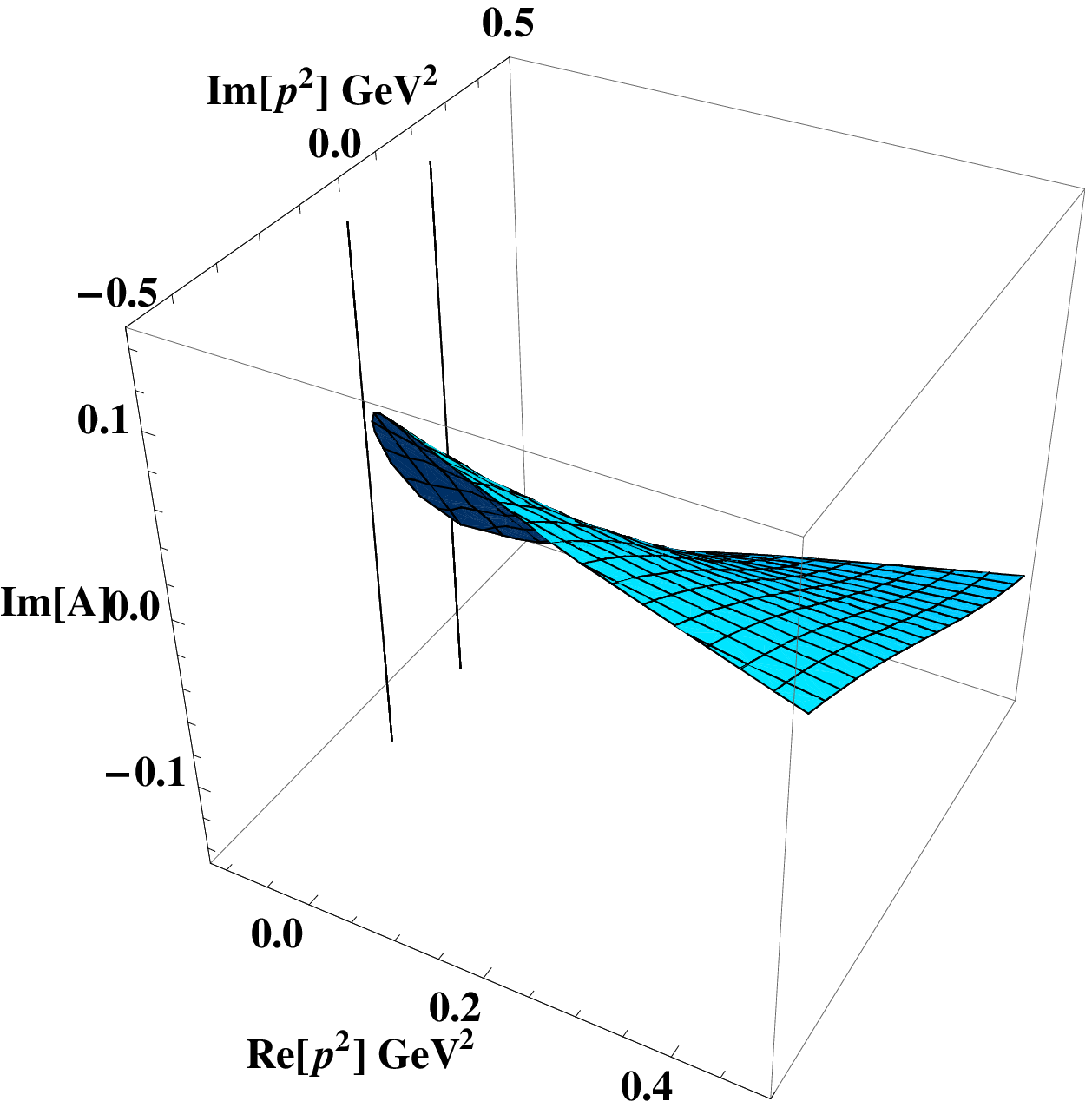,width=7cm}}
\centerline{\epsfig{file=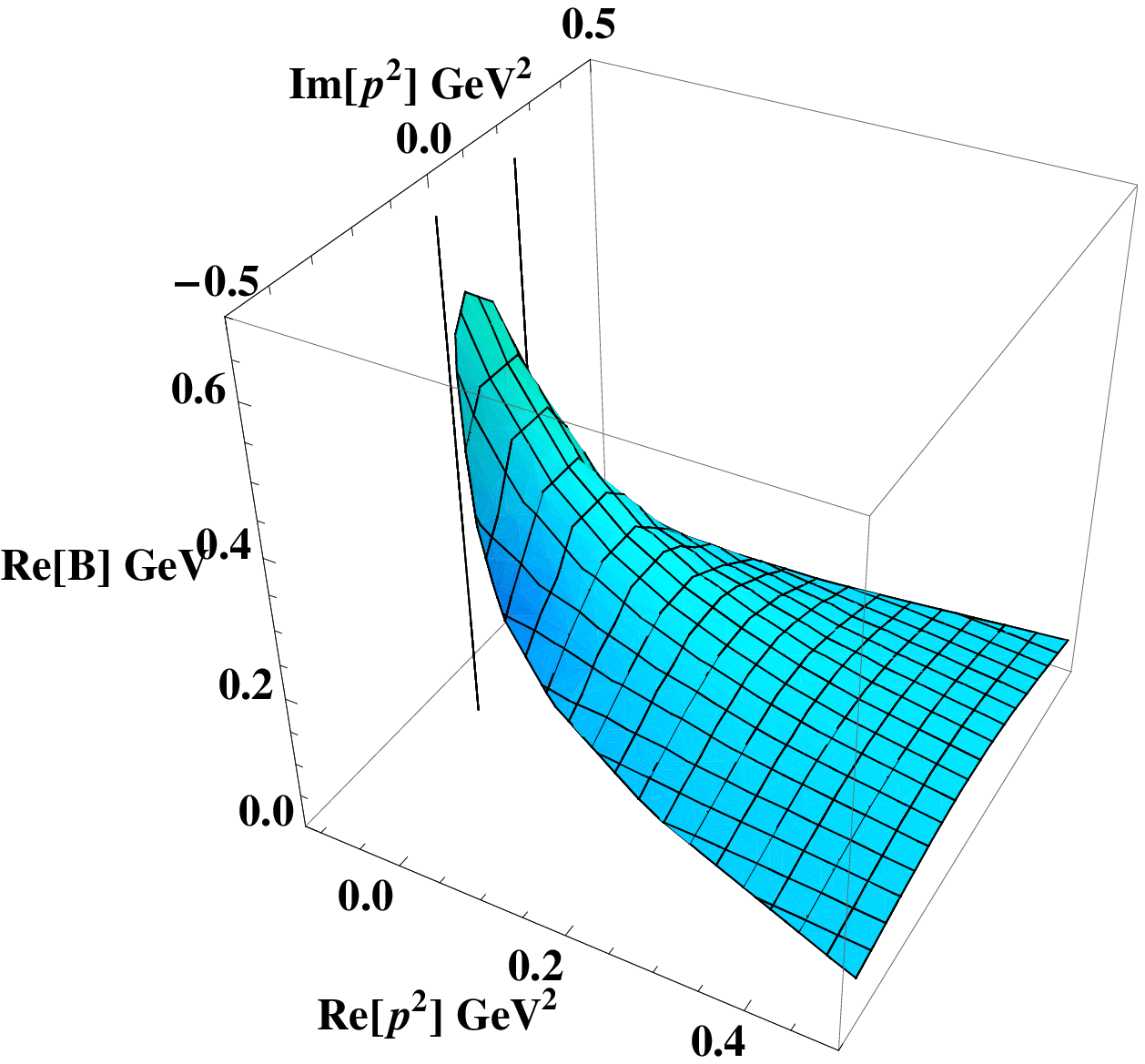,width=7cm}
		\epsfig{file=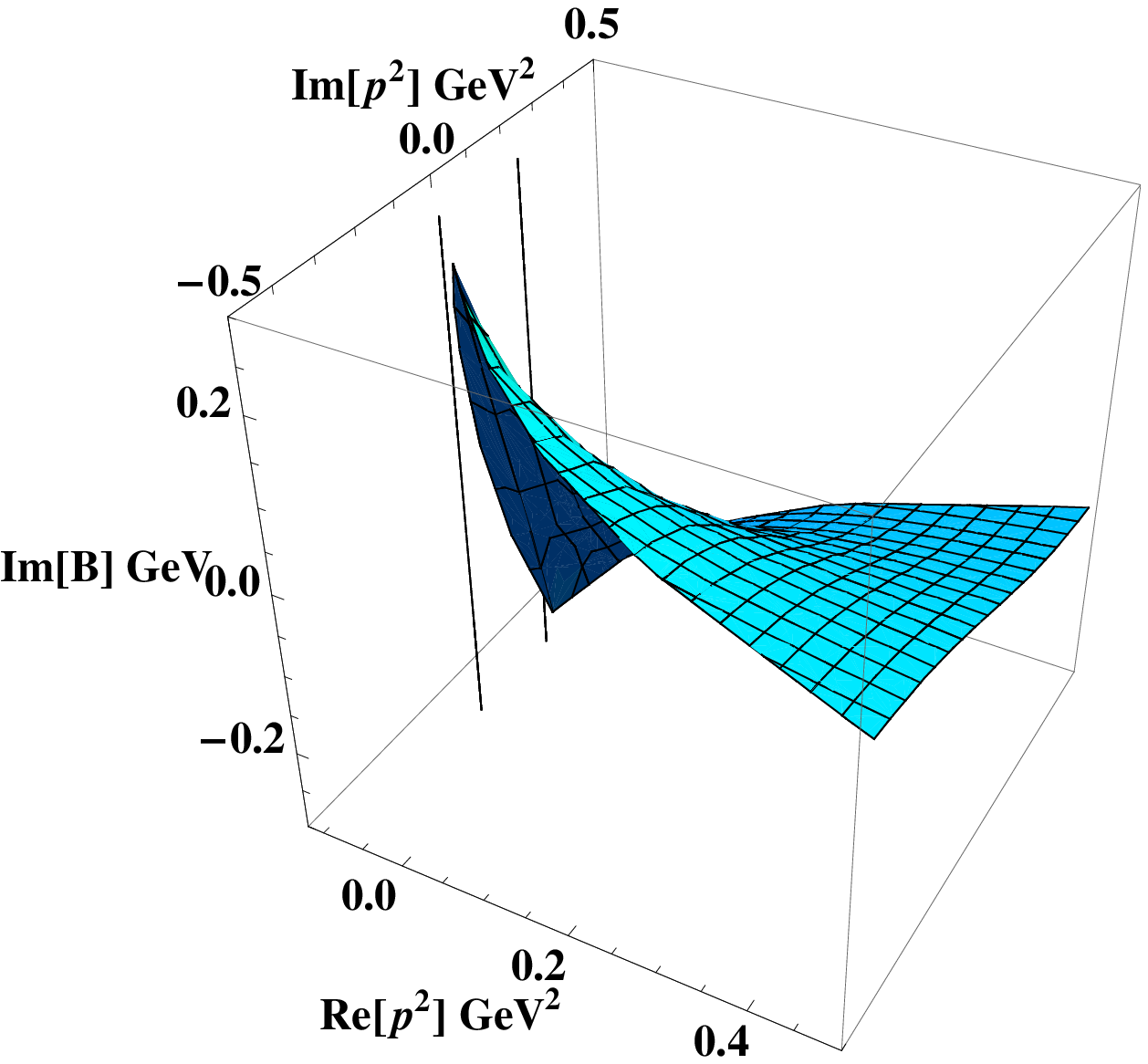,width=7cm}}
\caption{The real and imaginary parts of the quark self-energies $A$ and $B$
in the complex-plane for the 1BC vertex, including the
pion back-reaction. The vertical lines show the position of the complex
poles, here at $p^2=(-0.0626\pm i\, 0.111)\mathrm{GeV}^2$. Solutions without the
inclusion of pion unquenching effects are qualitatively similar.
\label{fig:cmplxsol}}
\end{figure}

\begin{table}[!t]
\begin{tabular}{|c||cc|}
\hline
	    & \multicolumn{2}{c|}{Vertex characterising the parabola $[\mathrm{GeV}^2]$} \\
          & \hspace*{1mm} Schwinger function \hspace*{0.5mm} 
	    & \hspace*{1mm} shell method      \hspace*{0.5mm}
	                                \rule[-2mm]{0mm}{5mm}\\ \hline\hline
1BC, m=0 MeV, wo $\pi$     & $(-0.087,0)$ & $(<-0.082,0)$ \rule[-2mm]{0mm}{5mm} \\\hline
1BC, m=3 MeV, wo $\pi$     & $(-0.104,0)$ & $(<-0.100,0)$ \rule[-2mm]{0mm}{5mm} \\\hline
1BC, m=0 MeV, with $\pi$   & $(-0.077,0)$ & $(<-0.072,0)$ \rule[-2mm]{0mm}{5mm} \\\hline
1BC, m=3 MeV, with $\pi$   & $(-0.092,0)$ & $(<-0.086,0)$ \rule[-2mm]{0mm}{5mm} \\\hline
\end{tabular}
\vspace{3mm}
\caption{The parabola on which the complex-conjugate poles are located, as
determined via the Schwinger-function and by the breakdown of the direct numerical procedure. The vertex, $(-m^2,0)$ is defined as the point at which the
parabola crosses the real-axis.\label{vertex-pole}}
\vspace{5mm}
\end{table}
To give a better picture on the influence of the pion back-reaction on the
dressing functions at complex momentum and also
to show the efficacy of the expanding shell method, detailed in
section~\ref{complex} and appendix~\ref{complexplane}, we show explicit 
solutions to the quark DSE in the complex-plane in
fig.~\ref{fig:cmplxsol}. For the purposes of demonstration, we employed
the 1BC vertex with a small quark mass $m(10\,{\rm GeV})=3\,{\rm MeV}$
and include the contribution from the pion back-reaction.

The advantages of employing such an expanding shell method and the
associated interpolation scheme becomes
apparent when we use our solutions in studies of the Bethe-Salpeter
equations. Not only are we able to provide solutions to the quark-DSE in
the complex plane for any \emph{numerically} determined gluon propagator or
quark-gluon vertex -- without changing or making assumptions of the
analytic structure by use of fit functions -- but we can do this quickly
and accurately. Indeed, it is gratifying to see that our bound-state
solutions change by less than a percent for a wide selection of coarse
and fine grids \cite{mesonpaper}.

As we expand our parabolas outward into the complex plane, our domain of
exploration approaches the location of the complex poles.  Close to
these poles, the function becomes steep and cannot be reliably
represented by our interpolation scheme without adaptive modification of 
the grid points. With such tuning, it is possible to see the onset of
conjugate poles in the complex plane by looking for bumps arising in the
solutions. Since we do not know the precise location of the pole or its
residue, we cannot extend our parabolas any further. Moreover, there
comes a point where the poles affect our numerical stability and lead to
a breakdown of the numerical procedure.

This breakdown allows us to determine
approximately on which parabola the complex poles are located, where
each parabola is characterised by its vertex $(-m^2,0)$, the point at
which the parabola crosses the real-axis. In table~\ref{vertex-pole} we list
the vertex of the parabolas on which the complex pole lies, as
determined from the Schwinger function and as inferred by the breakdown
of our numerical method.
Both methods are in excellent agreement. This provided additional justification
for the Fourier-transform method exploited in subsection \ref{complexresults}.

While this procedure can in principle be applied to the CP-vertex, whose
solutions contain a pole on the negative real-axis, perhaps accompanied
by a branch-cut, the numerical solution becomes much more involved due to the
derivative-like terms appearing in the vertex. By implementing a robust
numerical procedure that deals with this numerical singularities
correctly, we believe that our method is applicable to finding solutions
in the complex plane for such a vertex construction.
We defer a detailed calculation to future work.

\section{Conclusion}

We studied the analytic structure of the quark propagator with and
without the inclusion of pion effects, in order to compare and contrast
with Gribov's conjecture of quark confinement due to supercriticality 
of the colour charge. He advocated the viewpoint that these pions play 
an important r\^ole for the confinement of quarks, as indicated in the 
analytic structure of their propagator. Studying a truncation scheme 
that essentially includes all features introduced and studied by Gribov,
% , and employing fewer truncations in Landau 
% gauge, 
we determined the unquenching effects in the quark propagator
due to the back-reaction of pions onto the quarks. Our numerical results
agree nicely with corresponding lattice calculations thus underlining
the reliability of our truncation scheme. Investigating the analytic
structure of the quark propagator by means of its Schwinger functions 
and direct solutions of the quark-DSE in the complex plane we found 
that the inclusion of pion effects had no qualitative effect. 
This is the central result of our work: Gribov's conjecture does not 
seem to hold.

Instead, it is the relative strengths of the various tensor components 
that constitute the fully dressed quark-gluon vertex, in particular whether 
these are in agreement with those occurring in QED due to the Ward-identity. 
This finding is in agreement with a previous investigation of the structure 
of the quark propagator \cite{Alkofer:2003jj}. It relegates the question
of quark confinement due to positivity violations in its Schwinger function 
to a more refined determination of the details of the quark-gluon vertex,
see {\emph e.g.} \cite{Alkofer:2006gz}.

We also determined the multiple solutions of the quark-DSE as a function of 
the current quark mass. In the $M(p^2=0)-m$-plane we find a behaviour similar
to hysteresis effects in ferromagnets connected to the phenomenon of
dynamical symmetry breaking.
Close to the origin of this plane we find 
the critical behaviour of the Wigner solution to be connected to the form of 
the effective running coupling associated with the quark-gluon interaction.
Employing a sufficiently strong coupling with an infrared fixed point as also
being used in Gribov's work, we find a multitude of solutions near the trivial
point.

Our approximation scheme for the quark-gluon interaction as a composition of
a Yang-Mills part and a part due to the pion back-reaction onto the quark is
a modification of the one used in \cite{Fischer:2007ze}. This modification
leads to improved values for low energy constants as the chiral condensate and
the pion decay constant and therefore has the potential to describe pion cloud 
properties of mesons and baryons via bound state equations. This is further 
explored elsewhere \cite{mesonpaper}.

\section*{Acknowledgements}
This work has been supported by the Helmholtz-University 
Young Investigator Grant No VH-NG-332 and by the DFG under grant number 
Ni 1191/1-1.

\appendix

\section{Solving the Quark DSE in the Complex Plane \label{complexplane}}
\begin{figure}[t]
\centerline{\epsfig{file=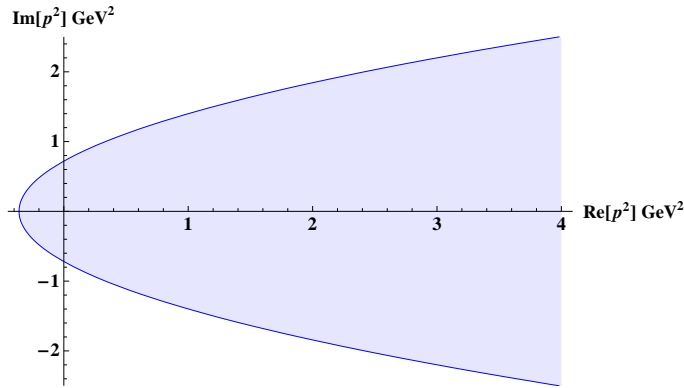,width=9cm}}
\caption{The bounded parabolic region in the $\mathbb{C}$-plane defined
by (\ref{eqn:app:momcmplx}), for equal momentum partitioning $\eta=1/2$, M =
$1.2$GeV. Solutions to the quark-DSE are required for the whole shaded
region, with $\textrm{Re}\left(p^2\right)$ extending as far as the UV cut-off.
}\label{fig:app:par1}
\end{figure}

Here we give details of our numerical method for calculating the quark
DSE for complex momenta.  In Euclidean space, our quark-DSE is
conventionally written as
\begin{equation}\label{eqn:app:dse}
S(p)^{-1} = Z_2 S_0^{-1}+g^2C_FZ_{1F}\int\frac{d^4q}{\left( 2\pi
\right)^4}\gamma_\mu S(q)\Gamma_\nu(q,k)D_{\mu\nu}(k)\;,
\end{equation}
with $k=p-q$ the momentum flowing through the gluon.
We want to solve this equation (in the previously described approximation) for
complex $p^2$ but using real $k^2$ only.
This also appears on consideration of
the Bethe-Salpeter equations. For those we find ourselves in need of the quark
propagator evaluated at momenta
\begin{equation}\label{eqn:app:momcmplx}
p_+ = p + \eta P\;,\qquad p_- = p+(1-\eta)P\;,
\end{equation}
with $\eta\in\left( 0,1 \right)$ a momentum partitioning parameter and $P$ the total momentum
of the meson. In Euclidean space, a bound state in the rest frame
has total momentum $P_4=iM$ and $\vec{P}=0$, and so the
momenta $p_{\pm}^2$ of (\ref{eqn:app:momcmplx}) define parabolic curves in
the complex plane. The necessary integrals over the angles $\left(p\cdot P\right)$ leads us to
require solutions to the quark-DSE for all complex momenta bounded by
these curves. For equal momentum partitioning, $\eta=1/2$, the region is
symmetric about the real-axis (see fig.~\ref{fig:app:par1} for an example).
The vertex of the parabola is located at $p^2 = \left( -M^2/4,0 \right)$ and
the focus at $p^2 = \left(0,0\right)$.

\begin{figure}[b]
\subfigure[]{
\epsfig{file=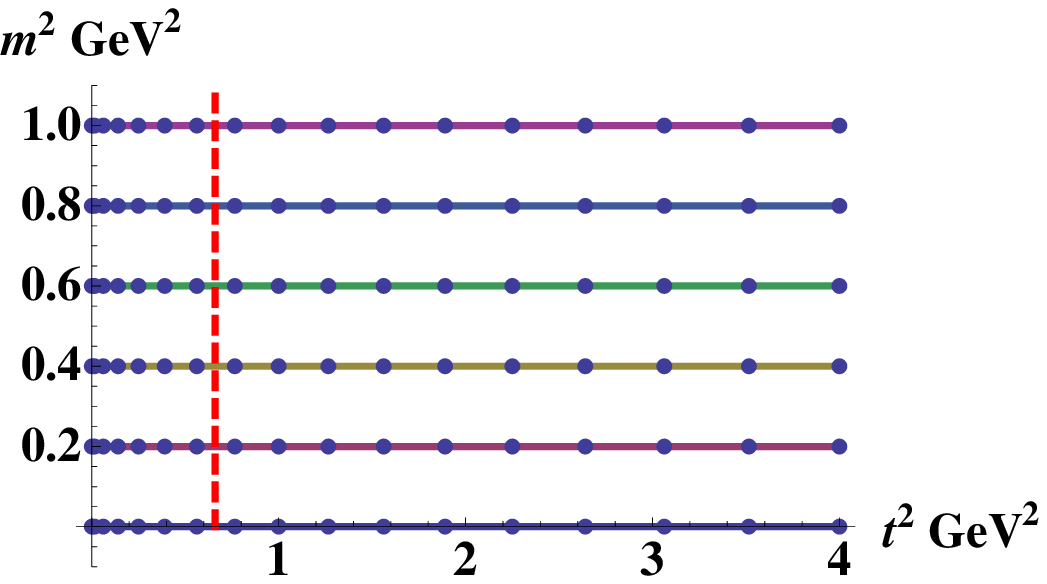,width=6.25cm}\label{app:fig:par2a}}
$\qquad
\stackrel{\mbox{\fontsize{24}{30}\selectfont
$\longrightarrow$}}{\textrm{mapping}}\qquad$
\subfigure[]{
\epsfig{file=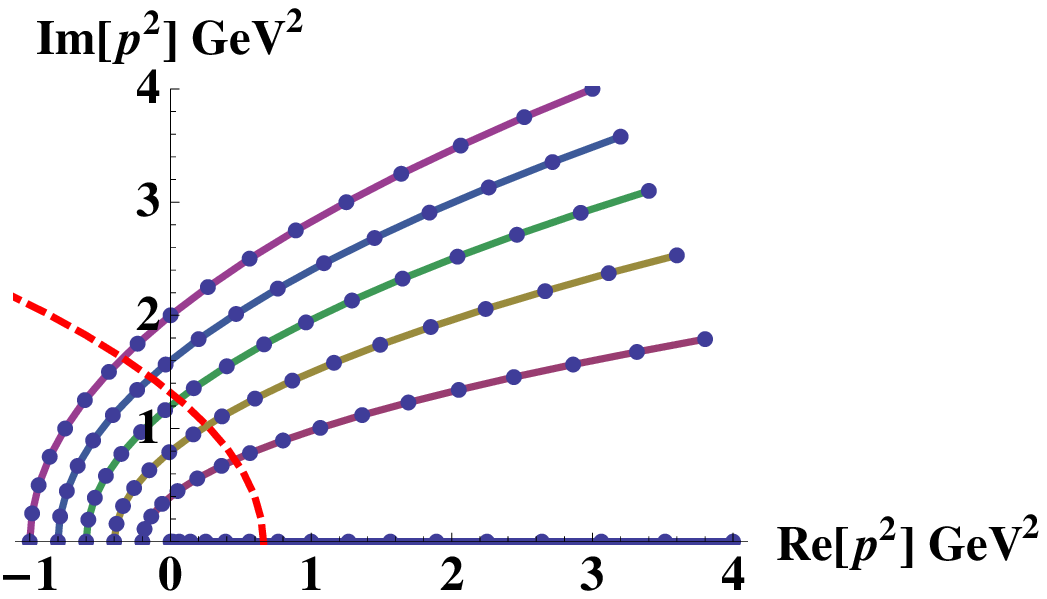,width=6.25cm}\label{app:fig:par2b}}
\caption{The mapping we employ to internally represent the parabolic
shells and the 2D interpolation}\label{app:fig:par2}
\end{figure}

In contemporary BS studies, one is generally forced to work in the
rainbow-ladder approximation whereby the full quark-gluon vertex
$\Gamma^\nu(q,k)$ of (\ref{eqn:app:dse}) is replaced by its bare
counterpart $\gamma^\nu$. We can then solve (\ref{eqn:app:dse}) for
$p^2\in\mathbb{C}$ without iteration, requiring only as input the quark propagator
on $\mathbb{R}^+$. In practice, we
need only perform calculations in the upper half-plane,
$\mathbb{H}^+$, since solutions are related by complex
conjugation. The caveat, however, is that we require knowledge of the
gluon interaction for complex momenta $k=p-q$. In general, though the
analytic structure of the gluon may be surmised, the analytic
continuation of some phenomenological ansatz is at best ill-defined.

To avoid evaluating the gluon for complex momenta, we modify the
momentum routing in the quark-DSE such that its momentum
is manifestly real. This amounts to introducing a shift
in the integration variable, $q\rightarrow k$, which is of
course valid in any translationally invariant regularisation scheme.
Such schemes are in general not employed in DSE studies, for
technical reasons, so one should be mindful of any boundary terms that
may arise; with a subtractive renormalisation scheme and careful consideration
of the renormalisation conditions these spurious terms may be rigorously eliminated. 
What remains is the equation:
\begin{equation}\label{eqn:app:dseshift}
S(p)^{-1} = Z_2 S_0^{-1}+g^2C_FZ_{1F}\int\frac{d^4k}{\left( 2\pi
\right)^4}\gamma^\mu S(q)\Gamma^\nu(q,k)D_{\mu\nu}(k)\;,
\end{equation}
where again we consider $p^2\in \mathbb{H}^+$ along the parabolas
of (\ref{eqn:app:momcmplx}), and now $k^2\in\mathbb{R}^+$,
$q^2\in\mathbb{C}$. Now that the integral equation depends on the
\emph{a priori} unknown quark propagator in the complex plane, we must
employ an iterative scheme to obtain solutions.

If we choose a point $p^2$ that lies on a parabola with vertex at
$(-m^2,0)$, then the integral equation only requires the quark propagator be
known in the region of the complex plane bounded by the \emph{same}
parabola. The most efficient way to obtain
solutions is then to expand in a series of parabolic shells stemming from the
real-axis, as shown in fig.~\ref{app:fig:par2b}.
To accelerate the
iteration process, the previously converged shell is extrapolated
outwards, using the Cauchy-Riemann equations, and used as an initial
guess for the next shell.

Because we are dealing with complex numbers, it is necessary to employ
some 2D interpolation scheme. Internally, our parabolic shells are
characterised by their vertex $m^2$ and a parameter $t^2$, shown as
stacks in
fig.~\ref{app:fig:par2a}, which are mapped onto the parabolas of
fig.~\ref{app:fig:par2b}. It is thus straightforward to take any point
$p^2\in\mathbb{C}$ and determine its corresponding value in $(t^2,m^2)$
space. Cubic-spline interpolation is used to interpolate along the
closest two shells $m^2_i\le m^2 < m^2_{i+1}$ in $t^2$, whilst linear interpolation in $\sqrt{m^2}$ is
sufficient for determining the value in-between. This essentially
leads to interpolation along a parabola, such as the dashed curve shown
in fig.~\ref{app:fig:par2}.

The drawback of this approach, however, is that without precise
information about the location of the poles and their residues, we are
unable to explore beyond the singularities appearing in the quark
propagator. This is discussed also in the main body of this work.

\end{document}